\documentclass[prb,a4paper,showpacs,twocolumn]{revtex4}

\usepackage{amssymb}
\usepackage{amsmath}
\usepackage{amsfonts}
\usepackage{graphicx}
\usepackage{bm}
\usepackage[usenames,dvipsnames]{color}

\renewcommand{\Re}{\,\textrm{Re}\,}
\renewcommand{\Im}{\,\textrm{Im}\,}
 
\DeclareMathOperator{\tr}{tr} 
\DeclareMathOperator{\sgn}{sgn} \DeclareMathOperator{\Imag}{Im}

\definecolor{DarkGreen}{rgb}{0.01,0.99,0.1}

 1

\begin{document}

\title{Relaxation dynamics of the electron distribution  in the Coulomb blockade problem}

\author{Ya.I.\,Rodionov$^{1}$,\ I.S.\,Burmistrov$^{1,2}$ and N.M.\,Chtchelkatchev$^{3,1,4}$ }

\affiliation{$^{1}$ L.D.\ Landau Institute for Theoretical Physics,
Russian Academy of Sciences, 117940 Moscow, Russia}
\affiliation{$^{2}$ Department of Theoretical Physics, Moscow
Institute of Physics and Technology, 141700 Moscow, Russia}
\affiliation{$^{3}$ Materials Science Division, Argonne National
Laboratory, Argonne, Illinois 60439, USA}

\affiliation{$^{4}$ Institute for High Pressure Physics, Russian
Academy of Sciences, Troitsk 142190, Moscow region, Russia}

\begin{abstract}
We study the relaxation dynamics of electron distribution function
on the island of a single electron transistor. We focus on the
regime of not very low temperatures in which an electron coherence
can be neglected but quantum fluctuations of charge are strong due
to Coulomb interaction. The quantum kinetic equation governing
evolution of the electron distribution function due to escape of
electrons to the reservoirs is derived. Analytical solutions for
time-dependence of the electron distribution are obtained in the
regimes of weak and strong Coulomb blockade. We find that usual
exponential in time relaxation is strongly modified due to the
presence of Coulomb interaction.
\end{abstract}
\date{\today}


\maketitle
\section{Introduction\label{Sec:Intro}}

Recently, the phenomenon of Coulomb blockade in single electron
devices~\cite{zaikin,ZPhys,grabert,aleiner,Glazman} has come into
the focus of the field of thermoelectricity.~\cite{Giazotto_review}
Among significant experimental achievements one can list development
of the Coulomb blockade thermometer,~\cite{Giazotto_review} the
thermal rectifier on the basis of a quantum dot,~\cite{ScheibnerNew}
new technique to measure temperature gradients across a quantum
dot,~\cite{Hoffmann} etc. The standard characteristic of
thermoelectric performance is the figure of merit which involves the
product of conductance, thermopower squared and inverse thermal
conductance. Measurements of thermopower and thermal conductance in
single electron transistors (SET) and quantum dots have been
performed during the last decade at different temperature
regimes.~\cite{Dzurak,Scheibner,Moeller} The theory of the
thermoelectric effects in electron devices has been put forward in
Refs.~[\onlinecite{Amman,BeenakkerStaring}]. During the last decade
the thermopower and thermal conductance have been studied in single
electron transistors and quantum
dots,~\cite{AndreevMatveev,TurekMatveev,MatveevAndreev1,kubala1,Nakanishi,Zianni,kubala2}
and in granular
metals~\cite{beloborodov0,TripathiLoh,beloborodov1,beloborodov2,beloborodov3}
in various regimes. However, the thermopower and thermal conductance
are linear response parameters and, therefore, describe the
equilibrium properties of a system only.

   In contrast, our work is focused on properties of single electron
devices in the out-of-equilibrium regime  which has attracted a lot
of theoretical interest recently.
In particular, the conductance of a quantum dot under ac pumping in
the stationary non-equilibrium state was obtained in
Ref.~[\onlinecite{BaskoKravtsov}], the current noise of an ac-biased
quantum dot was studied in Ref.~[\onlinecite{Bagrets}], the
non-equilibrium dephasing rate and zero-bias anomaly in single
electron transistor was computed in
Ref.~[\onlinecite{AltlandEgger}], the statistics of temperature and
current fluctuations in the fully out-of equilibrium single electron
transistor was investigated in
Refs.~[\onlinecite{nazarov1,nazarov2}], and the extension of the
$P(E)$-theory~\cite{NazPE} to the out-of-equilibrium regime has been
developed in Refs.~[\onlinecite{K1,K2}]. However, these works dealt
with regimes when the distribution function of electrons in a
quantum dot or an island of single electron transistor were fixed by
external sources, e.g., ac or dc bias voltage.

In this paper we address a different question: how does an electron
distribution function once being prepared relaxes toward the
equilibrium state in the Coulomb blockade problem. Apart from
general physical interest in understanding of a non-equilibrium
regime, the answer to this question is important for the field of
electron thermometry.~\cite{Giazotto_review}

\begin{figure}[t]
  \includegraphics[width=50mm]{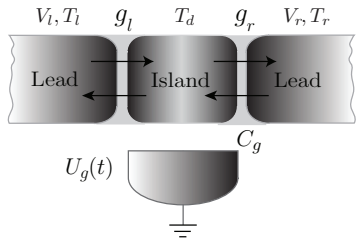}
  \caption{\label{figure1}
    The sketch of a SET device. The leads are kept at different
    temperatures (chemical potentials) inducing
    heat (electric) currents.
          }
\end{figure}

We consider the simplest system: single electron transistor. The
set-up is shown schematically in Fig.~\ref{figure1}. Metallic island
is coupled to an equilibrium electron reservoirs via tunneling
junctions. Depending on the task, the reservoirs and the island may
be kept at different temperatures\ ($T_l$, $T_r$ respectively),
different chemical potentials (constant or varying in time:
$\mu_l(t)$, $\mu_r(t)$) or  quasi-stationary gate voltage
($U_g(t)=U_0+U_\omega\cos\omega t$) may be applied to the system.
The physics of the system is governed by several energy scales: the
Thouless energy of an island\ $E_{\rm Th}$, the charging energy\
$E_c$, and the mean single-particle level spacing\ $\delta$.
Throughout the paper the Thouless energy is considered to be the
largest scale in the problem. This allows us to treat the metallic
island as a zero dimensional object with vanishing internal
resistance. The dimensionless total conductance (in units $e^2/h$)
of tunneling junctions\ $g$\ is an essential control parameter. The
junctions are assumed to have a large number of channels but the
conductance of each one is assumed to be small $g^{(l,r)}_{\rm
ch}\ll1$. The temperature is assumed to be low enough: $T\ll
\max\{1,g\}E_c$ in order to keep electrons strongly correlated due
to Coulomb interaction. At low temperatures the interplay of Coulomb
interaction and electron coherence dominates the physics of single
electron devices. Account of both effects is a formidable
undertaking. Therefore, we are going to restrict ourselves to the
regime of not very low temperatures in which an electron coherence
can be neglected but quantum fluctuations of charge are strong due
to Coulomb interaction. Then, the physics of the system is
adequately described in the framework of Ambegaokar-Eckern-Sch\"{o}n
(AES) effective action.~\cite{ambegaokar} The effective action
describes the system in terms of bosonic field\ $\varphi$, which is
usually termed as the {\it plasmon} field. Its time derivative is
interpreted as fluctuating electric potential of electrons inside
the island.
 AES-approach has well-known limitations. Deriving AES
action one assumes that the products of electron Green's functions
averaged over disorder are substituted with products of
disorder-averaged Green's functions in every calculation. That is
why the processes of phase-coherent multiple impurity scattering
inside the island are left out. The limitations\ in the regime\
$g\gg1$ and $g\ll1$ were discussed in detail in
Refs.~[\onlinecite{efetov-bel}] and
~[\onlinecite{Efetov-Tschersich}], respectively. It was shown that
at temperatures\ $T\gg \max\{1,g\}\delta$, AES-action approach is
justified. Following Ref.~[\onlinecite{AltlandMeyer}], we shall term
the temperature range, $\max\{1,g\} E_c\gg T\gg \max\{1,g\}\delta$
as an {\it interaction without coherence} regime.   This
`interaction without coherence' regime is an attainable experimental
reality, e.g. in experiments reported in Refs.~[\onlinecite{Dzurak}]
and~[\onlinecite{Scheibner}] the necesary conditions were satisfied.

In the case of strong Coulomb blockade ($g\ll1$), the theoretical
study of relaxation of an electron distribution in the interaction
without coherence regime has been done before for a single quantum
dot [\onlinecite{beloborodov4}] and for an 1D array of quantum dots
[\onlinecite{beloborodov-glatz}]. However, the considerations of
Ref.~[\onlinecite{beloborodov4}] have been restricted by assumptions
that i) the electron distribution is the Fermi function with some
temperature different from the equilibrium one; ii) transport is
dominated by co-tunneling processes (Coulomb valley regime); iii)
temperatures of the island and the reservoirs are close to each
other.

In the present paper, we undertake the analysis of relaxation of an
electron distribution function which is free of above-mentioned
restrictions.

Since we are going to capture non-equlibrium physics, we employ the
formalism of AES-action in its out-of-equilibrium form throughout
the paper. We supplement it with quantum kinetic equation to explore
relaxation dynamics of electron distribution. For a SET with large
number of tunneling channels we derive the quantum kinetic equation
with the collision integral due to escape of electrons to the
reservoirs. It is valid in the entire span of values of\ $g$ and
generalizes the  one obtained in Ref.[\onlinecite{BaskoKravtsov}]
for sequential tunneling (first order in $g$) and cotunneling
(second order in $g$) approximations in the framework of the
orthodox theory of the Coulomb blockade. In fact, our collision
integral is always an infinite series in powers of $g$. Indeed, each
tunneling event is accompanied by the radiation of a plasmon. That
is why the collision integral becomes of the infinite order in the
distribution function of electrons inside the island. This situation
is entirely different from the  one in Fermi liquid and leads to
non-trivial relaxation.

As a test of the quantum kinetic equation, in the regime of linear
response we derive analytical expressions for transport
coefficients: conductance, thermal conductance and the response of
electric current to temperature difference.  In the regime of weak
Coulomb blockade ($g\gg1$) we establish the following new results
for the transport coefficients: i) the conductance and thermal
conductance violate Wiedemann-Franz law,  and deviation of the
Lorentz ratio\ $\mathcal{L}$ from value $\pi^2/3e^2$ demonstrates
weak periodic dependence on the gate voltage; ii) the thermopower
weakly oscillates with the gate voltage around zero value. Weak
oscillations of the Lorentz ratio and thermopower with the gate
voltage found in the regime $g\gg 1$ are manifestation of the known
gate-voltage dependence of these
quantities~\cite{Amman,BeenakkerStaring,AndreevMatveev,
TurekMatveev,MatveevAndreev1,kubala1,Nakanishi,Zianni,kubala2} in
the strong Coulomb blockade regime, $g\ll 1$.

In weak and strong Coulomb blockade regimes we have employed the
quantum kinetic equation to solve the relaxation of the electron
distribution in two cases: i) the distribution of electrons inside
the island is the Fermi-function with some temperature; ii) the
distribution function of electrons inside the island is arbitrary.
In the former case we have managed to extract the relaxation
dynamics of the electron temperature; in the latter case we have
obtained evolution of a distribution function itself. In both cases
we assumed that electron escape to reservoirs is the primary
relaxation mechanism. In general, the collision integral in the
quantum kinetic equation is non-local in energy due to inelastic
nature of tunneling processes: the radiation of plasmon always
accompanies the tunneling event. In a number of wide parametric
regimes: weak Coulomb blockade and Coulomb peak in the strong
Coulomb blockade, the kernel of the quantum kinetic equation
acquires a quasi-elastic form. However, the collision integral
remains non-local in energy due to renormalization effects in these
cases. The co-tunneling regime is qualitatively different: the
kernel of the collision integral is entirely inelastic.

Our new result is that despite quasi-elastic form of the collision
integral, strong Coulomb interaction dramatically changes the
relaxation laws comparing to simple exponential ones expected from
golden-rule type arguments. They suggest that electron relaxation
rate is to be proportional to the width of electrons' levels inside
the island, $g\delta$, prompting simple exponential relaxation. The
renormalization effects due to Coulomb interaction make the width of
electrons' levels dependent on the electron distribution and lead to
the non-exponential relaxation laws. For example, in the regime of
the sequential tunneling, we have discovered that there is a time
regime in which relaxation of the electron temperature in a SET
island is independent of the tunneling conductance $g$.

The paper is organized as follows. In Sec.~\ref{FORMALISM} we
introduce the hamiltonian and essential parameters of the problem.
Sec.~\ref{KINETIC} is devoted to the out-of-equilibrium AES-model
and to derivation of the quantum kinetic equation. Sec.
\ref{TRANSPORT} is devoted to derivation of general expressions for
the linear response coefficients. The relaxation dynamics of
electrons in the island is explored  in the weak ($g\gg1$) and
strong ($g\ll1$) Coulomb blockade regimes in
Sec.~\ref{RELAX-WEAK}-\ref{RELAX-STRONG}. Discussion of the results,
 comparision with other relaxation mechanisms, different from
electron escape to reservoirs and conclusions are presented in
Sec.~\ref{DISCUSSION}.


\section{Formalism\label{FORMALISM}}

A SET is described by the Hamiltonian
\begin{gather}
   \label{ham1}
      H=H_0+H_c+H_t ,
\end{gather}
where
\begin{gather}
   \label{ham2}
    H_0=\sum_{k,i}\varepsilon^{(i)}_{k}a^{(i)\dagger}_{k}a^{(i)}_{k}+
    \sum_\alpha\varepsilon^{(d)}_\alpha d^\dagger_{\alpha}d_{\alpha}.
\end{gather}
describes free electrons in the leads and the island,\
$H_c$\ describes Coulomb interaction of carriers in the island, and\
$H_t$\ describes the tunneling.
Here operators $a^{(i)\dag}_{k}$ ($d^\dag_{\alpha}$)\ create a
carrier in the $i$-th lead (island). Then, the tunneling hamiltonian is
\begin{gather}
   \label{ham-tun}
    H_{t}=\sum_{k,\alpha,i}t^{(i)}_{k\alpha} a_{k}^{(i)\dagger} d_{\alpha}+{\rm
    H.c.}
\end{gather}
The charging Hamiltonian of electrons in the box is taken in the
capacitive form:
\begin{gather}
   \label{ham3}
      H_c=E_c\big(\hat{n}_d-q\big)^2.
\end{gather}
Here $E_c= e^2/(2C)$\ denotes the charging energy, and\
$\hat{n}_d$\ is an operator of a particle number in the island:
\begin{gather}
   \label{number}
    \begin{split}
    \hat{n}_d=\sum_{\alpha}d^\dagger_{\alpha}d_{\alpha}.
    \end{split}
\end{gather}

To characterize the tunneling it is convenient to introduce the
following hermitean matrices:
\begin{gather}
   \hat{g}^{(i)}_{kk^\prime}=(2\pi)^2
   \left[\delta(\varepsilon^{(i)}_{k})\delta(\varepsilon^{(i)}_{k^\prime})\right]^{1/2}
   \sum_\alpha t_{k\alpha}
   \delta(\varepsilon^{(d)}_{\alpha})t^\dagger_{\alpha
   k^\prime},\label{gkk}\\
   \check{{g}}^{(i)}_{\alpha\alpha^\prime}=(2\pi)^2
   \left[\delta(\varepsilon^{(d)}_{\alpha})\delta(\varepsilon^{(d)}_{\alpha^\prime})
   \right]^{1/2}\sum_k t^\dagger_{\alpha
   k}
   \delta(\varepsilon^{(i)}_{k})t_{k\alpha^\prime},\label{gaa}
\end{gather}
The first of
them acting in the Hilbert space of the states of the lead, the
second -- in the space of the islands states. The energies
$\varepsilon^{(i)}_k,\varepsilon^{(d)}_\alpha$ are accounted from the Fermi
level, and the delta-functions should be smoothed on the scale
$\delta E$, such that $\max\{\delta,\delta^{(l,r)}\} \ll\delta E\ll T$.
Here, $\delta$ and $\delta^{(l,r)}$ stand for mean level spacing of
single-particle states on the island and reservoirs, respectively.
The classical dimensionless conductance (in units $e^2/h$) of the
junction between a reservoir and the island can be  expressed as follows~\cite{Glazman}
\begin{gather}
 \label{conductance-def1}
 g=g_l+g_r,\qquad g_{l,r}=\sum_k \hat{g}^{(l,r)}_{kk}
 \equiv \sum_\alpha\check{g}^{(l,r)}_{\alpha\alpha} .
\end{gather}
Therefore, each non-zero eigenvalue of $\hat g^{(i)}$ or $\check g^{(i)}$
corresponds to the transmittance of some `transport' channel between a reservoir
and the island.~\cite{landauer} The effective number of these `transport' channels
($N^{(i)}_{\rm ch}$) is given by
\begin{equation}
N^{(i)}_{\rm ch} = \frac{\left ( \sum\limits_k \hat{g}^{(i)}_{kk} \right )^2}
{\sum\limits_{kk^\prime} \hat g^{(i)}_{kk^\prime}\hat g^{(i)}_{k^\prime k}}
\equiv \frac{\left ( \sum\limits_\alpha \check{g}^{(i)}_{\alpha\alpha} \right )^2}
{\sum\limits_{\alpha\alpha^\prime} \hat g^{(i)}_{\alpha\alpha^\prime}
\hat g^{(i)}_{\alpha^\prime \alpha}} .
\end{equation}
The effective dimensionless conductance $g^{(i)}_{\rm ch}$ of a `transport' channel
can be written as follows
\begin{gather}
\label{aes-condition1}
g^{(i)}_{\rm ch}=  \frac{\sum\limits_{kk^\prime}
\hat g^{(i)}_{kk^\prime}\hat g^{(i)}_{k^\prime k}}{\sum\limits_k \hat{g}^{(i)}_{kk} }
\equiv \frac{\sum\limits_{\alpha\alpha^\prime} \hat g^{(i)}_{\alpha\alpha^\prime}
\hat g^{(i)}_{\alpha^\prime \alpha}}{\sum\limits_\alpha \check{g}^{(i)}_{\alpha\alpha}} .
\end{gather}
The dimensionless conductance $g$ then becomes
\begin{equation}
g=g_l+g_r, \qquad g_{l,r} = g^{(l,r)}_{\rm ch}N^{(l,r)}_{\rm ch} .
\end{equation}
In what follows we will always assume
\begin{gather}
\label{aes-condition2}
g^{(i)}_{\rm ch}\ll 1, \qquad N^{(i)}_{\rm ch}\gg 1 .
\end{gather}
Notice that under these circumstances the conductances $g_{l,r}$ can still
be large provided the effective number of channels $N^{(l,r)}_{\rm ch}$ is
sufficiently large.

\section{Action and kinetic equaitons\label{KINETIC}}

\subsection{AES-action}
To tackle the system  which is out of equilibrium we have to employ
essentially non-equilibrium formalism. Keldysh technique is thus the
only way through. We employ Keldysh form of AES-action (we sketch
the known details of derivation in Appendix
\ref{APPENDIX-AES}):~\cite{zaikin,Skvortsov}
\begin{gather}
   S=S_{c}+S_d \label{S_AES}
\end{gather}
where
\begin{gather}
   S_{c}=\frac{1}{E_c}\int\dot{\varphi_c}\dot{\varphi_q}dt+2q \int\dot{\varphi}_q dt . \label{S_C_AES}
\end{gather}
Here $\varphi_{c,q}=(\varphi_+\pm\varphi_-)/2$ with $\varphi_\pm$
denoting bosonic field on both branches of Keldysh contour.
Physically, the bosonic field is associated with the  fluctuating
electric potential on the island. In terms of classic and quantum
boson exponents
\begin{gather}
    X_{c,q}=\frac{1}{\sqrt{2}}\big(e^{i\varphi_+}\pm e^{i\varphi_-}\big) , \label{Xs}
\end{gather}
the dissipative part of AES-action reads:
\begin{gather}
\label{aes1}
 \begin{split}
   S_d&=\frac{g}{4}\int \Big(\bar{X}_c(t)\bar{X}_q(t)\Big)\\
       &\times\begin{pmatrix}
         0 &  \Pi^A(t,t^\prime)\\
        \Pi^R(t,t^\prime) &\ \Pi^K(t,t^\prime)
       \end{pmatrix}
       \begin{pmatrix}
             X_c(t^\prime)\\
             X_q(t^\prime)
         \end{pmatrix}dtdt^\prime.
 \end{split}
\end{gather}
Here $\Pi^{R,A,K}$\ are corresponding components of electron
polarization operator in the Keldysh space. They are given by a standard
formulae presented for reference in
Appendix \ref{APPENDIX-AES}. In a case of constant density of states (DOS) in
the island and leads the kernel of the AES action can be simplified:\
\begin{gather}
  \label{aes2}
   \Pi^{R,A,K}(t,t^\prime)=\int\frac{d\omega}{2\pi}\Pi^{R,A,K}_\omega(\tau)
   e^{-i\omega(t-t^\prime)} ,\\
   \Pi^{R,A}_\omega(\tau)= \mp i\sum_\alpha\frac{g_\alpha}{g}\int
   \big[F^d_\varepsilon(\tau)-F^\alpha_{\varepsilon-\omega}(\tau)\big]
   \frac{d\varepsilon}{2\pi},\label{aes2a} \\
        \Pi^K_\omega(\tau)= 2i \sum_\alpha\frac{g_\alpha}{g}\int(1-F^d_\varepsilon(\tau)
        F^\alpha_{\varepsilon-\omega}(\tau))\frac{d\varepsilon}{2\pi}.\label{aes2b}
\end{gather}
Here we define a slow time $\tau=(t+t^\prime)/2$. Function\
$F_\varepsilon(\tau)$\ is given in terms of the Wigner transform
$f_\varepsilon(\tau)$ of the  electron distribution function\
$f(t,t^\prime)$: $F_\varepsilon(\tau)=1-2f_\varepsilon(\tau)$.
$\Pi^{R,A,K}_\omega(\tau)$ are  Wigner transforms of corresponding
functions in time domain.

As seen from the structure of the r.h.s. of \eqref{aes2b}, it is
suitable to introduce function\ $F^r_\varepsilon=(\sum_\alpha
g_\alpha F^\alpha_\varepsilon)/g$ which may be called the effective
distribution function of two reservoirs. It is that combination of
reservoir distribution functions that enters all the following
equations of the paper.

Although Eq.~\eqref{aes2a} is exact we neglect all derivatives with
respect to slow time $\tau$ in Eq.~\eqref{aes2b}. It is also
convenient to introduce function\ $B_\omega(\tau)$\ in accordance
with
\begin{gather}
  \label{BP}
 \Pi^K_\omega(\tau)=2i\Imag\Pi^R_\omega(\tau) B_\omega(\tau)
\end{gather}
relating Keldysh and retarded (advanced) components of polarization
operator. The function $B_\omega(\tau)$ plays a role of a
distributiion function for electron-hole excitations. In the
equilibrium it is given by $\coth(\omega/2T)$.

In what follows we assume that electrons in the leads are locally
thermalized such that $f_\varepsilon^{l,r}(\tau)$ are the
Fermi-functions. Depending on the parameters of the model, both
quasi-equilibrium and non-equilibrium regimes can exist in the
island. Therefore, we assume $F_\varepsilon^d(\tau)$ to be an
arbitrary function slow varying with time $\tau$, and derive the
kinetic equation for $F_\varepsilon^d(\tau)$ by which the AES action
should be supplemented.

\subsection{Kinetic equations}
The starting point for deriving kinetic equation for a SET is the
Dyson equation for the Keldysh component of electron's Green's
function:~\cite{KEStandard}
\begin{equation}
 \label{kinetic0}
    (\partial_t+\partial_{t^\prime})F^{d}(t,t^\prime) =
  \frac{i}{4\pi^2\nu_d} \Bigl [\Sigma^K-\Sigma^R \cdot F^d + F^d \cdot
  \Sigma^A\Bigr ]_{t,t^\prime}.
\end{equation}
Here, $\nu_d= \delta^{-1}=\overline{\sum_\alpha
\delta(\varepsilon^{d}_\alpha)}$ is an averaged single particle
density of states in the island  and $\Sigma^{K,R,A}$ are the
components of self-energy in Keldysh space.
 To the second-order in tunneling Hamiltonian $H_T$ (lowest order
in $1/N_{\rm ch}$) the Wigner transform of the self-energy shown in
Fig.~\ref{figure2}a reads (see Appendix~\ref{APPENDIX-2})
\begin{figure}[t]
  \includegraphics[width=85mm]{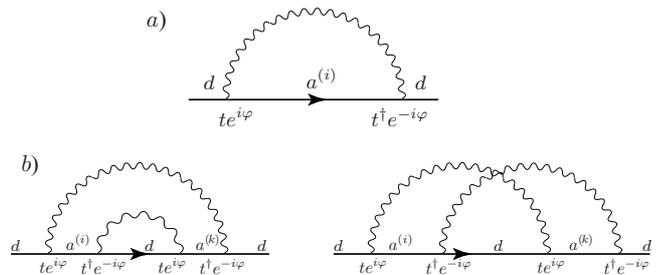}
  \caption{\label{figure2}
     Feynman diagrams for fermion self-energy: a) second order in $H_T$;
     b) fourth order in $H_T$.
}
\end{figure}
\begin{gather}
  \Sigma^{R,A}_{\varepsilon}(\tau) = \pm\frac{\pi g}{2} \int\frac{d\omega}{2\pi}\Bigl [
 {\cal D}^K_\omega(\tau)\pm 2 F^r_{\varepsilon-\omega}(\tau){\cal D}^{R,A}_\omega(\tau)\Bigr ] ,\notag \\
  \Sigma^{K}_{\varepsilon}(\tau) = \pi g \int\frac{d\omega}{2\pi}\Bigl [
  F^r_{\varepsilon-\omega}(\tau) {\cal D}^K_\omega(\tau) +2i\Imag{\cal
  D}^{R}_\omega(\tau)\Bigr ] . \label{self-energies2}
\end{gather}
Here we perform Wigner transform of the exact sel-energies and introduce the correlation functions of boson exponents:
\begin{gather}
 \begin{split}
  {\cal D}^R(t,t^\prime)&=-i\langle
  X_c(t)\bar{X}_q(t^\prime)\rangle ,\\
  {\cal D}^A(t,t^\prime)&=-i\langle
  X_q(t)\bar{X}_c(t^\prime)\rangle ,\\
  {\cal D}^K(t,t^\prime)&=-i\langle X_c(t)\bar{X}_c(t^\prime)\rangle .
 \end{split}
\end{gather}
It is convenient to parametrize ${\cal D}^K(t,t^\prime)$ via the boson distribution
function $\mathcal{B}(t,t^\prime)$:
\begin{gather}
    {\cal D}^K_{t,t^\prime}=\Big({\cal D}^R\cdot{\cal B}-{\cal B}\cdot{\cal D}^A\Big)_{t,t^\prime} .
\end{gather}
 It is worthwhile to mention that the next ( fourth
order in $H_T$) contribution to the self-energy which is shown in
Fig.~\ref{figure2}b is of the order $g^2/N_{ch}$. This correction to
the self-energy is of the same order as terms omitted in the course
of derivation of the AES action~\eqref{aes1}. In the considered
limit $N_{ch} \gg 1$ it can be safely neglected.

Performing Wigner transform of Eq.~\eqref{kinetic0} and neglecting
all slow time derivatives in its r.h.s. we obtain the quantum kinetic equation for the distribution
function of electrons on the island of a SET:
\begin{gather}
 \label{ke-set}
   \begin{split}
  &\partial_\tau F^{d}_{\varepsilon}(\tau) = -\sum_{\alpha=l,r}\frac{g_\alpha}{2\pi\nu_d}
  \int\frac{d\omega}{2\pi}
  \Im{\cal D}^{R}_\omega(\tau)\\
  &\times \Bigl\{ \Bigl (
  F^\alpha_{\varepsilon-\omega}(\tau)-F^d_{\varepsilon}(\tau)\Bigr )
  {\cal B}_\omega(\tau) +1 -F^d_{\varepsilon}(\tau)F^\alpha_{\varepsilon-\omega}(\tau)\Bigr\}  .
   \end{split}
\end{gather}
This quantum kinetic equation constitutes one of the main results of the present paper.
It describes evolution of the distribution function $F_\varepsilon^d(\tau)$ of electrons in the island
due to interaction with boson field $\varphi$ and tunneling to the leads and back. The quantum
kinetic equation~\eqref{ke-set} is derived for any values of $g_r$ and $g_l$; the r.h.s.
of Eq.~\eqref{ke-set} can be written as the series in powers of $g$ due to the presence
of $\Im {\cal D}_\omega^R(\tau)$ and ${\cal B}_\omega$.
The boson distribution ${\cal B}_\omega$ is determined by electron distribution
function $F_\varepsilon^d(\tau)$ and should be found from the solution of the AES-action~\eqref{S_AES}.

At $g\gg 1$ the kernel ($\Im {\cal D}^R_\omega$) of the collision
integral of the quantum kinetic equation~\eqref{ke-set} resembles
the kernel of the collision integral in the quantum kinetic equation
for disordered electron liquid~\cite{Schmid,AA,AA1} for energy
transfers $\omega \gg g \delta$ as it is expected.~\cite{efetov-bel}
At $g\ll 1$ the quantum kinetic equation~\eqref{ke-set}  which takes
into account the renormalization effects via $\Im {\cal D}^R_\omega$
generalizes the kinetic equation derived in
Ref.~[\onlinecite{BaskoKravtsov}] in the framework of the orthodox
theory~\cite{kulik} for sequential tunneling and inelastic
cotunneling approximations.


\section{Transport coefficients\label{TRANSPORT}}

Using quantum kinetic equation~\eqref{ke-set} we are able to derive general
formulae for all linear response coefficients of the SET for any value of
 $g$. Voltage $(\Delta V = V_r-V_l)$ and
temperature $(\Delta T=T_r-T_l)$ differences across the SET cause charge
$(I^{(e)})$ and heat $(I^{(q)})$\ currents. Electric and thermoelectric transport
coefficients are defined as
\begin{gather}
    \begin{pmatrix}
          I^{(e)}\\
          I^{(q)}
    \end{pmatrix}=
    \begin{pmatrix}
          G_V & G_T\\
          M   & K
    \end{pmatrix}
    \begin{pmatrix}
          \Delta V\\
          \Delta T
    \end{pmatrix}.
\end{gather}
Here coefficients\ $M$ and $G_T$ (the response of a heat current to
voltage difference and the response of electric current to
temperature difference, respectively)  are related via Onsager
relation $M=G_T T$.~\cite{Abrikosov} The thermal conductance
$\kappa$ is usually defined as $\kappa=K-G_V T S^2$ where
$S=G_T/G_V$\ stands for the thermopower. The electric and heat
currents in the $\alpha$-th reservoir can be found as
\begin{gather}
{}\,   \hspace{-2cm} \begin{pmatrix}
          I^{(e)}_\alpha\\
          I^{(q)}_\alpha
    \end{pmatrix}=-\frac{g_\alpha}{4\pi}\int d\varepsilon
    \begin{pmatrix}
          1\\
          \varepsilon
    \end{pmatrix}
  \int\frac{d\omega}{2\pi}\Im{\cal D}^{R}_\omega(\tau) \\
 \times  \Bigl \{ \Bigl
   (F^d_{\varepsilon+\omega}(\tau)-F^\alpha_{\varepsilon}(\tau)\Bigr )
   {\cal B}_\omega(\tau)
   -1+F^\alpha_{\varepsilon}(\tau)F^d_{\varepsilon+\omega}(\tau)\Bigr\}. \notag
\end{gather}
The current conservation corresponds to the condition
$I^e_l+I^e_r=0$. It fixes the boson distribution function ${\cal
B}_\omega$ to be equal to the electron-hole distribution function
$B_\omega$ introduced in Eq.~\eqref{BP}:
\begin{eqnarray}
B_\omega(\tau) &=& \frac{\Pi_\omega^K(\tau)}{2i\Im \Pi_\omega^R(\tau)} \notag \\
&=& \frac{\sum\limits_{\alpha=l,r}g_\alpha\int d\varepsilon \Bigl [
   1 -F^d_{\varepsilon}(\tau)F^\alpha_{\varepsilon-\omega}(\tau)\Bigr]}
   {\sum\limits_{\alpha=l,r}g_\alpha\int d\varepsilon \Bigl [
    F^d_{\varepsilon}(\tau)-F^\alpha_{\varepsilon-\omega}(\tau)\Bigr ]}. \label{BO1}
\end{eqnarray}

The heat current conservation  $I^q_l+I^q_r=0$\ determines the equilibrium
temperature of the island:
\begin{gather}
  \label{T-eq}
   T^{\rm(eq)}_{d}=\frac{g_lT_l+g_rT_r}{g_l+g_r}.
\end{gather}
A straightforward computation of charge and heat currents gives
\begin{gather}
  \begin{split}
    \begin{pmatrix}
          I^{(e)}_l\\
          I^{(q)}_l
    \end{pmatrix}&=-\frac{g_lg_r}{g}\frac{e}{4\pi}\int\frac{d\omega}{2\pi}
     \frac{\Imag{\cal D^R_\omega}}{\sinh^2\frac{\beta\omega}{2}}\\
     &\times\begin{pmatrix}
          \beta\omega & -\frac{(\beta\omega)^2}{2}\\
            -\frac{\beta\omega^2}{2}  & \omega\frac{\pi^2+(\beta\omega)^2}{3}
    \end{pmatrix}
    \begin{pmatrix}
          e\Delta V\\
          \Delta T
    \end{pmatrix} .
  \end{split}
\end{gather}

Introducing the quantities $g^\prime, g_T^\prime$, and $k^\prime$ as
\begin{gather}
  \label{set-g}
  \begin{split}
  \begin{pmatrix}
          G_V & G_T\\
          M   &  K
    \end{pmatrix} &= \frac{e^2}{h}\frac{g_lg_r}{(g_l+g_r)^2}
    \begin{pmatrix}
          g^\prime & -\frac{1}{e}g_T^\prime\\
          -\frac{T}{e}g_T^\prime &  \frac{T}{e^2}k^\prime
    \end{pmatrix} ,
  \end{split}
\end{gather}
we obtain
\begin{gather}
  \label{response0}
   \begin{pmatrix}
          g^\prime\\
          g_T^\prime\\
          k^\prime
    \end{pmatrix}=-g\int\frac{d\omega}{4\pi}\frac{\Imag{\cal
    D^R_\omega}}{\sinh^2\frac{\beta\omega}{2}}\beta\omega
    \begin{pmatrix}
          1\\
          \frac{\beta\omega}{2}\\
          \frac{\pi^2+(\beta\omega)^2}{3}
    \end{pmatrix} .
\end{gather}
 We stress that Eq.~\eqref{response0} is  valid
for any value of tunneling conductance\ $g$. It generalizes
expressions for transport coefficients obtained in
Refs.~[\onlinecite{schoeller,kubala2}] for $g\ll 1$ to arbitrary
values of $g$.

It is worthwhile to express Eq.~\eqref{response0} in terms of the
tunneling density of states of electrons in the island (see
Appendix~\ref{AppTDOS}):
\begin{gather}
  \label{dos}
   \nu_d(\varepsilon)=-\nu_d\int\Im
   {\cal D}^R_\omega\Bigl\{ \coth\frac{\omega}{2T}-\tanh\frac{\varepsilon+\omega}{2T}\Bigr\}\frac{d\omega}{2\pi}.
\end{gather}

  Substituting expression \eqref{dos} for the
 tunneling density of states and performing standard integrals
 with Fermi and Bose distribution functions one can check that the
 results \eqref{response0} are allowed to be {\it exactly} rewritten in the form
\begin{gather}
   \label{fermi1}
   \begin{pmatrix}
          g^\prime\\
          g_T^\prime\\
          k^\prime
    \end{pmatrix}=
    g\int d\varepsilon\frac{\nu_d(\varepsilon)}{\nu_d}\bigg(-\frac{\partial
    f_\varepsilon^d}{\partial\varepsilon}\bigg)
    \begin{pmatrix}
          1\\
          \varepsilon\\
          \varepsilon^2
    \end{pmatrix} .
\end{gather}
Eq.~\eqref{fermi1} for the transport coefficients
resembles the corresponding expression in the
Fermi-liquid.~\cite{AGD,Abrikosov} However, contrary to
Fermi-liquid, the tunneling density of states $\nu_d(\varepsilon)$
has strong dependence on energy for $ \varepsilon\to 0$. In general,
$\nu_d(\varepsilon) =  \nu_d^{\rm even}(\varepsilon)+\nu_d^{\rm
odd}(\varepsilon)$ where $\nu_d^{\rm even/odd}(\varepsilon)$ is
even/odd function of $\varepsilon$. It can be shown that $\nu_d^{\rm
even/odd}(\varepsilon)$ is even/odd function of the external charge
$q$. Therefore, $g^\prime$ and $k^\prime$ are even functions of $q$
whereas $g_T^\prime$ is an odd function of the external charge.

For macroscopic samples of ordinary metals, the Wiedemann-Franz law
provides a universal relation between the conductance and thermal
conductance. It states that the Lorenz ratio $\mathcal{L} =
\kappa/(G_V T)$ , is a constant given by the Lorenz number
$\pi^2/3e^2$. As follows from Eq.~\eqref{fermi1}, one can expect the
violation of Wiedemann-Franz law in the presence of strong
dependence of $\nu_d(\varepsilon)$ on electron energy.

 In the case $g\gg 1$ one is able to perform
perturbative expansion in $1/g$ and take into account
non-perturbative corrections. The function $\Im {\cal D}_\omega^R$
acquires the following form in the equilibrium~\cite{burmistrov2}
\begin{gather}
\Im {\cal D}^R_\omega =- \frac{\pi}{T} \left [1- \frac{2}{g}\ln
 \frac{gE_ce^\gamma}{2\pi^2 T} - \frac{g^2 E_c}{\pi^2 T} e^{-g/2} \cos2\pi q \right ]\omega
 \delta(\omega) \notag\\
 - \frac{2\pi}{g \omega} \left [ 1 + \frac{g^3 E_c}{2 \pi^2 T} e^{-g/2}
 \cos 2\pi q \right ]  \notag \\
 + \frac{g^2 E_c}{\pi T } e^{-g/2} \cos 2\pi q \frac{\omega}{\omega^2+4\pi^2 T^2} \notag \\
 - \frac{g^2 E_c}{T } e^{-g/2} \sin 2\pi q \left ( \delta(\omega) - \frac{2 T}
 {\omega^2+4\pi^2 T^2} \right ) . \label{ImDR_eq}
\end{gather}
Here, function $\omega\delta(\omega)$ can be understood as $\Im
a/[\pi(\omega+a+i0)]$ where the limit $a\to 0$ should be performed
at the very end of all calculations, (this calculation can be, e.g.
integration over $\omega$).
The non-perturbative in $1/g$ corrections (exponential terms
$\exp(-g/2)$) come from Korshunov instantons~\cite{korshunov} of the
AES-action. Then by using Eq.~\eqref{ImDR_eq} we find from
Eq.~\eqref{response0}
\begin{gather}
 \label{response1}
   g^\prime=g-2\ln\frac{gE_c}{T}-\frac{g^3E_c}{6T}e^{-g/2}\cos2\pi q,\\
   g^\prime_T=-\frac{2g^3E_c}{\pi T}e^{-g/2}\Big(1-\frac{\pi^2}{12}\Big)
   \sin2\pi q, \label{response1a}\\
   k^\prime=\frac{\pi^2}{3}\Big(g^\prime+\frac{4}{3}+\frac{2g^3E_c}{\pi^2T}\cos2\pi q
   \Big[\frac{\pi^2}{3}-3\Big]\Big). \label{response1b}
\end{gather}
The result for $g^\prime$ has been obtained in
Ref.~[\onlinecite{AltlandMeyer}].
Equations~\eqref{response1a}-\eqref{response1b} are new and valid
for temperatures $T \gg g^2 E_c \exp(-g/2)$. We emphasize that
$g_T^\prime$ has only non-perturbative instanton) contribution. The
same holds for the thermopower:
\begin{equation}\label{Sfin}
S = - \frac{2g^2 E_c}{\pi e T} \left (1- \frac{\pi^2}{12}\right )
\exp \left ( -\frac{g}{2}\right ) \sin 2\pi q .
\end{equation}
At $g\gg 1$ the violation of the Wiedemann-Franz law is weak and the Lorentz number is given as
\begin{equation}\label{Lfin}
\mathcal{L} = \frac{\pi^2}{3 e^2} \Biggl [ 1 + \frac{4}{3g} +
\frac{2}{3}\left (1-\frac{9}{\pi^2}\right ) \frac{g^2 E_c}{T}
e^{-g/2} \cos 2\pi q \Biggr ] .
\end{equation}
Due to the presence of the non-perturbative contribution, the
Lorentz number is temperature dependent and oscillates as a function
of the external charge $q$. Eqs.~\eqref{Sfin} and \eqref{Lfin}
constitute one of the main results of the present paper.

 In the strong coupling regime\ $g\ll 1$, Eq.~\eqref{response0} supplemented
 by the proper expression for $\Im {\cal D}^R_\omega$
 (cf. Eqs.~\eqref{final1} and \eqref{cot1}) results in exactly the same expressions for
 the transport coefficients as obtained in Refs.~[\onlinecite{schoeller,kubala1, kubala2,
 Zianni}]. We refer a reader to these works for details.


\section{Relaxation of electrons in the island, weak coupling regime $g\gg1$\label{RELAX-WEAK}}

Next, we want to illustrate the ability of quantum kinetic
equation~\eqref{ke-set}  combined with fine field-theoretical
scaling of essential physical quantities. We consider the problem of
relaxation of electrons in the island towards the equilibrium due to
the tunneling to the reservoirs and back. There are two possible
scenarios. The first one can be refered to as a quasi-equilibrium
regime. The electron distribution inside the island is given by the
Fermi-function but with non-equilibrium temperature $T_d$ which
slowly relaxes to its equilibrium value. The second scenario is
fully non-equilibrium regime when electron distribution is
arbitrary. Which scenario persists depends on the ratio
$\tau_E/\tau_{ee}$ where
$\tau_E$ stands for
the energy relaxation time due to tunneling mechanism and
$\tau_{ee}$ for the energy relaxation time due to electron-electron
interaction in the island. The non-equilibrium regime persists
provided $\tau_E \ll \tau_{ee}$ and the quasi-equilibrium regime is
possible if $\tau_E \gg \tau_{ee}$. We will argue below (see
Sec.~\ref{DISCUSSION}) that both scenarios are possible.

 There is one more relaxation time involved:
$\tau_{RC}$ which determines relaxation of the electric charge on the island.
 In the weak Coulomb blockade regime, $\tau_{RC}$ is given by the following
 clasical estimate: $\tau_{RC} \simeq 2\pi/gE_c$. As we shall see below,
 $\tau_E \gg \tau_{RC}$. Therefore, it is allowed to assume that at first
 there is quick relaxation of the electric charge on the island and, then,
 slow relaxation of the electron distribution function or temperature
 towards the equilibrium. Technically, it means that initial electron distribution
 function $F_\varepsilon^d(0)$ satisfies the constraint
 $\int d\varepsilon [F_\varepsilon^d(0)-F_\varepsilon^r] =0$.

 As
was discussed in the Introduction, the
 renormalization of physical observables drastically
changes the relaxation dynamics of the system. Therefore, before
solving kinetic equation we need to establish the scaling of a
theory's coupling constants under non-equilibrium conditions.

\subsection{Renormalization of AES-action at $g\gg1$}

The AES-action is renormalized due to its nonlinear form. In the
equilibrium case renormalization of the action  is well-known (see
e.g., Ref.~[\onlinecite{zaikin}]). In our case non-equilibrium makes
the problem non-trivial. As in equilibrium, we expect the necessary
scaling of the coupling constant\ $g$. The additional question that
inevitably arises is whether the structure of the kernel of AES
action\ (the components of polarization operator $\Pi_\omega$ in
Keldysh space in Eq.~\eqref{aes1})\ is changed due to
renormalization? The details of the calculation are presented in
Appendix~\ref{RENORM}. We prove that the structure of the bare
action is fully restored, the kernel of the AES action being intact
during renormalization group (RG) procedure. The coupling constant
renormalizes according to
\begin{gather}
    \label{rg1}
 g(\underline{\Lambda})=g(\overline{\Lambda})
   - 2 \int^{\overline{\Lambda}}_{\omega=\underline{\Lambda}}
   \frac{B_\omega(\tau)}{\omega}d\omega .
\end{gather}
Here high energy scale $\overline \Lambda$ is naturally set by the
first term in Eq.~\eqref{S_C_AES}: $\overline \Lambda \sim g E_c$.
To demonstrate that the integral in Eq.~\eqref{rg1} is indeed
logarithmic we explore the behavior of the integrand at\
$\omega\rightarrow\infty$. It is straightforward to get the
following asymptotic for function $B_\omega$ at $\omega\to\infty$:
\begin{gather}
   \label{b-equation3}
   \begin{split}
   B_\omega&=\sgn\omega+\delta B_\omega,\\
   \delta B_\omega&=(F^d_\omega-\sgn\omega)\\
   &+\frac{1}{2\omega}\sum_\alpha\frac{g_\alpha}{g}
   \int(\varepsilon+\omega)\big(F^d_{\varepsilon+\omega}-F^d_{\omega}\big)
   \partial_\varepsilon F^\alpha_\varepsilon d\varepsilon .
  \end{split}
\end{gather}
 We expect that any physical distribution function obeys the
condition $F^{d}_\varepsilon\rightarrow\sgn\,\varepsilon$ at
$\varepsilon\rightarrow\infty$. Then
\begin{gather}
   \lim_{\omega\rightarrow\infty} \delta B_\omega=0 .
\end{gather}
Therefore, the high-energy asymptotic of function $B_\omega$ is
given by $\sgn \varepsilon$ as in the equilibrium. This way, the
logarithmic behavior of integral in Eq.~\eqref{rg1} is ascertained.

To get the renormalized action one has to integrate out all the
frequencies down to the lowest scale\ $\omega_0$, at which the RG
stops.This energy scale can be determined as
\begin{gather}
   \delta B(\omega_0)\sim 1.
\end{gather}
Let $\varepsilon_d$\ be a characteristic energy scale of the island
distribution function (the scale at which electron distribution
function\ $F^d_\varepsilon$ becomes almost equal to $\sgn
\varepsilon$). Then one can easily check that the following estimate
holds (see Appendix~\ref{RENORM} for elaborate details)
\begin{gather}
   \omega_0 \sim\max\{\varepsilon_d,T_r,T_l\}, \label{Cond_O0}
\end{gather}
where\ $T_r,T_l$\ are temperatures of the reservoirs.
Energy scale $\omega_0$
serves as a natural lower cut-off, $\underline{\Lambda}=\omega_0$,
in the RG procedure. Finally, we find
\begin{equation}
g(\omega_0) = g - 2 \ln \frac{g E_c}{\omega_0}.\label{Cond_O1}
\end{equation}
In the equilibrium, $\varepsilon_d=T_r=T_l=T$ and one finds
$\omega_0 =T$. Eqs.~\eqref{Cond_O0}-\eqref{Cond_O1}
describe renormalization of the AES-action under non-equilibrium
conditions.

\subsection{Non-equilibrium regime}

The relaxation problem is formulated as follows. At $t=0$ the island
is heated and some electron distribution function $F_\varepsilon^d(0)$ is
created. The characteristic energy $\varepsilon_d$ of electrons in the
island is larger than temperatures of the reservoirs, which are kept
fixed and equal to each other\ $\varepsilon_d>T_r=T_l$. The system is
released and the island is cooling down due to the tunneling of
electrons to the reservoirs and back.

Performing expansion to the second order in boson fields $\varphi_{c,q}$,
one straightforwardly finds (see Appendix~\ref{RENORM})
\begin{gather}
  \label{im1}
    \Imag {\cal D}^R_\omega=-\frac{2\pi\delta(\omega)}{
    B_\omega}\Big(1- \frac{1}{g}\int \frac{B_{\omega}}{\omega}d\omega \Big)
    -\frac{2\pi}{g \omega}.
\end{gather}

We mention that this result generalizes the
perturbative (independent of $q$) part of Eq.~\eqref{ImDR_eq} to the
non-equilibrium case. With the help of~\eqref{im1} one can compute
the collision integral in the r.h.s. of the quantum kinetic
equation~\eqref{ke-set} and obtain
\begin{gather}
  \begin{split}
    \partial_\tau F_\varepsilon^d(\tau)&=-\frac{ \mathcal{G}(\tau)\delta}{2\pi}(F^d_\varepsilon(\tau)-F^r_\varepsilon),\\
     {\cal G}(\tau)&=g-\int\frac{B_\omega(\tau)}{\omega}\,d\omega.
  \end{split} \label{KEsim1}
\end{gather}
 Here we neglect last term in Eq.~\eqref{im1} for the following reasons. It gives
 contribution
of the order of unity whereas the first term in Eq.~\eqref{im1} involves
$\int d\omega\,B_\omega/\omega \sim \ln g E_c/\varepsilon_d \gg 1$.
 Although, Eq.~\eqref{KEsim1} has a quasi-elastic form, in fact, it is
highly non-linear equation: $\mathcal{G}(\tau)$ involves information
about the electron distribution at all energies.

As was shown above, the quantity\ ${\cal
G}(\tau)$\ has meaning of the renormalized coupling constant of the theory. Simple algebra
leads us to the differential equation for
the function $\mathcal{G}(\tau)$:~\cite{Footnote1}
\begin{gather}
 \label{diff1}
    \partial_\tau {\cal G}(\tau)=-\frac{\delta{\cal
    G}(\tau)}{2\pi}\Big({\cal G}(\tau)-{\cal G}_r\Big) ,
\\    {\cal G}_r=g-\int\frac{d\omega}{\omega}\coth\frac{\omega}{2T_r}  .\notag
\end{gather}
The solution reads
\begin{gather}
   {\cal G}(\tau)=\frac{\mathcal{G}(0) \mathcal{G}_r}
   {\mathcal{G}(0)+\big(\mathcal{G}_r- \mathcal{G}(0)\big)
   e^{-\frac{\mathcal{G}_r\delta}{2\pi}\tau}} .
\end{gather}
Now by using this result we integrate Eq.~\eqref{KEsim1} and obtain the evolution of the electron
distribution function\ $F^d_\varepsilon$:
\begin{eqnarray}
    F^d_\varepsilon(\tau)&=&F_\varepsilon^r + \Bigl (F^d_\varepsilon(0)-F_\varepsilon^r\Bigr )\notag \\
    &\times &\left \{ \frac{\mathcal{G}(0)}{\mathcal{G}_r}
    \left [\exp\left (\frac{\mathcal{G}_r\delta}{2\pi}\tau\right )-1\right ]+1 \right \}^{-1}. \label{EqFf}
\end{eqnarray}

 Eq.~\eqref{EqFf} demonstrates energetically
uniform relaxation of the electron distribution function. This fact
is a direct consequence of the quasi-elastic form of the kinetic
equation~\eqref{KEsim1}. However, due to renormalization effects the
form of the relaxation law is different from the exponential one.

Let us define the characteristic energy $\varepsilon_d$ as
$\varepsilon_d^2-T_r^2 = (3/\pi^2)\int d\varepsilon\, \varepsilon
(F_\varepsilon^r-F_\varepsilon^d)$ such that $\varepsilon_d=T_d$ in
the quasi-equilibrium case $F_\varepsilon^d  = \tanh
(\varepsilon/2T_d)$. Then, in the case $\varepsilon_d(0) \gg T_r$
and at not too long times $\tau \ll 2\pi/\delta \mathcal{G}_r$, one
finds from Eq.~\eqref{EqFf} that the characteristic energy decreases
according to the power-law:
\begin{equation}
\varepsilon_d(\tau) = \varepsilon_d(0)
\left [1 + \frac{\delta\mathcal{G}(0)}{2\pi}\tau \right ]^{-1/2}. \label{Ed}
\end{equation}

\subsection{Quasi-equilibrium regime}

In the quasi-equilibrium regime, we need to take into account the collision
integral $I^{(ee)}_{\varepsilon}$ due to electron-electron interaction in the
island.~\cite{AA,Schmid}
As this term is added to the r.h.s. of Eq.~\eqref{ke-set}, it makes the electron
distribution $F_\varepsilon^d$ to
be the Fermi-function. Multiplying both parts of Eq.~\eqref{ke-set} by $\varepsilon$
and integrating them
over energy, we obtain the following equation
(the well-known identity $\int d\varepsilon\, \varepsilon I^{(ee)}_{\varepsilon} = 0$ is used):
\begin{gather}
   \label{relax1}
   \frac{d T^2_d}{d\tau}=-\frac{g\delta}{2\pi}(T_d^2-T_r^2) .
\end{gather}
Here we use the leading (classical) part
of Eq.~\eqref{im1} ($\Im \mathcal{D}^R_\omega=-2\pi\delta(\omega)/B_\omega$).
Equation~\eqref{relax1} yields standard exponential relaxation towards the equilibrium. In
the limit $T_d\gg T_r$ ($\mathcal{G}(0)\gg \mathcal{G}_r$) it is also possible to compute
collision
integral using the entire one-loop expression~\eqref{im1} of its
kernel. Naturally, one-loop correction reveals itself in the
logarithmic renormalization of $g$ in the r.h.s. of Eq.~\eqref{relax1}. By using Eq.~\eqref{im1},
we find
\begin{gather}
   \label{relax2}
   \begin{split}
   \frac{d T^2_d}{d\tau}&=-\frac{\mathcal{G}(\tau)\delta}{2\pi}T_d^2,\\
                {\cal G}(\tau)&=g-2\ln\frac{gE_c}{c T_d(\tau)}
   \end{split}
\end{gather}
where $c$ is a numerical constant of the order of unity which does
not influence final results. The solution of~\eqref{relax2} reads:
\begin{gather}
   \label{sol-relax1}
   \begin{split}
    T_d(\tau)&=T_d(0)\exp\bigg(-\frac{{\cal
    G}(0)}{2}\Big[1-e^{-\delta\tau/2\pi}\Big]\bigg)\\
    \tau&\ll\frac{2\pi}{\delta}\ln\frac{\mathcal{G}(0)}{\mathcal{G}_r}.
   \end{split}
\end{gather}
The condition in the second line of Eq.~\eqref{sol-relax1} implies
that solution holds for not too long times at which $T_d(\tau)\gg
T_r$ (${\cal G}(\tau)\gg{\cal G}_r$). 
The logarithmic
renormalization of the conductance changes the character of
temperature relaxation. At long times $2\pi/\delta \ll \tau \ll
(2\pi/\delta)\ln\mathcal{G}(0)/\mathcal{G}_r$, the cooling of the
island slows down in comparison to the standard exponential decay
which is developed at short times $\tau\ll 2\pi/\delta$:
\begin{gather}
  T_d(\tau)=T_d(0)e^{-\frac{{\cal G}(0) \delta \tau}{4\pi}} . \label{Td}
\end{gather}

 It is instructive to compare the relaxation of temperature in
the quasi-equilibrium regime
and the characteristic energy $\varepsilon_d$ in the non-equilibrium regime
given by Eqs.~\eqref{Td} and \eqref{Ed} at times $\tau \ll 2\pi/\delta \mathcal{G}_r$,
respectively. While the former demonstrates exponential behavior, the latter decreases
in accordance with the power-law.


\section{Relaxation of electrons in the island, strong coupling regime,\ $g\ll1$\label{RELAX-STRONG}}

In the strong coupling regime there are two possible scenarios for
relaxation of electrons in the island of the SET. The first one
persists if
 $\tau_E \ll \tau_{ee}$. In this non-equilibrium case
the carriers inside the island do not have time to thermalize and to
form the Fermi-distribution with some temperature. In this
case the the time evolution of distribution function itself becomes
the main objective. This task is solved in section~\ref{EDrel} below.
The second scenario develops in the opposite limit, $\tau_E\gg \tau_{ee}$. Namely, the
relaxation rate due to electron-electron interaction inside the
island is much faster than the rate due to electron tunneling
through the contacts. Thus, the temperature of carriers in the
island  becomes a well defined characteristic of a system.
Consequently, the relaxation of the island's temperature will be the
focus of our analysis in section~\ref{Trel}.

 As in the previous section we shall assume that the
electric charge on the island quickly relaxes and only then slow
relaxation of the electron distribution or temperature starts. In
the strong Coulomb blockade regime this picture is well justified
since $1/\tau_{RC} \simeq g \max\{T,\Delta\} \gg 1/\tau_E$.

\begin{figure}[t]
   \includegraphics[width=65mm]{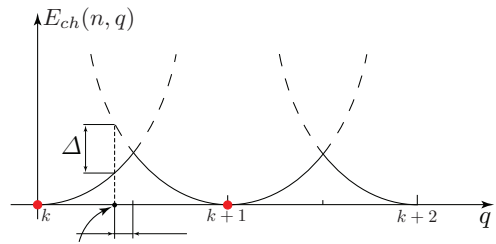}
    \caption{\label{figure3}
    (Color online) Charging energy\ $E_{ch}=E_c(n-q)^2$\ as a function of gate charge $q$.
          }
\end{figure}

We concentrate on
the most interesting case: the vicinity of a degeneracy point:\
$q=k+1/2$\ where $k$ is an integer. Following
Ref.~[\onlinecite{matveev}], the hamiltonian
\eqref{ham1}-\eqref{ham3} can be simplified by truncating the
Hilbert space of electrons on the island to two charging states:
with\ $Q=k$\ and\ $Q=k+1$ (see Fig.~\ref{figure3}). The projected
hamiltonian then takes a form of $2\times2$\ matrix acting in the
space of these two charging states. Denoting the deviation of the
external charge from the degeneracy point by $\Delta$:
$q=k+1/2-\Delta/(2E_c)$\ we write the projected hamiltonian
as:~\cite{matveev}
\begin{gather}
  \label{ham4}
   H=H_0 + H_t + \Delta S_z + \frac{\Delta^2}{4E_c}+\frac{E_c}{4}
\end{gather}
where $H_0$ is given by Eq.~\eqref{ham2} and
\begin{equation}
  \label{ham5}
 H_t=\sum_{k,\alpha}t_{k\alpha} a^\dagger_k d_\alpha
   S^-+\hbox{h.c.}
\end{equation}
Here $S^z,\ S^{\pm}=S^x\pm iS^y$\ are ordinary (iso)spin\ $1/2$\
operators.
\subsection{Non-equilibrium pseudo-fermions}
To deal with spin operators it is standard to use Abrikosov's
pseudo-fermion technique.~\cite{abrikosov} We introduce
two-component pseudo-fermion operators $\psi^\dagger_\alpha$,
$\psi_\alpha$ such that
\begin{gather}
   \label{pseudofermion}
  S^i=\psi^\dagger_\alpha S^i_{\alpha\beta}\psi_\beta .
\end{gather}
The out-of-equilibrium pseudo-fermions were tackled before.~\cite{wingreen,woelfle} As usual,
one introduces Keldysh contour,
doubling the number of fermions. The system is out of equilibrium
and one has to be very cautious.  The distribution function\ ${\cal
F}_{\varepsilon}$\ of pseudo-fermions is not known a priori. Rather, it is
to be defined self-consistently from corresponding kinetic
equation. Pseudo-fermions are also subject to constraint on their
number:
\begin{gather}
       \mathcal{N}(t)=\sum_\alpha\psi_\alpha^\dagger(t)\psi_\alpha(t)=1
\end{gather}
Thus the state of a system ought to be projected on the state with\
$\mathcal{N}=1$\ at any instant of time. The operator of particle number is
conserved by Hamiltonian~\eqref{ham4}-\eqref{ham5}. Consequently,
the operator of projection on to physical subspace\ $\mathcal{N}=1$\ commutes
with hamiltonian too. It means that the projection on to physical
subspace is needed at a single point of Keldysh contour only. We
insert the factor\
$\exp (\eta \sum_\alpha\bar{\psi}_\alpha\psi_\alpha )$\ into density
matrix and take the limit\ $\eta\rightarrow-\infty$ at the end of
any diagrammatic calculation. Then
\begin{gather}
 \label{pf-partition0}
  \langle{\cal O}\rangle=\lim_{\eta\rightarrow-\infty}\frac{\langle{\cal O N}\rangle_{\rm pf}}{\langle{\cal N}\rangle_{\rm pf}}  .
\end{gather}
Provided the operator
$\mathcal{O}$ has zero expectation value in the sector with zero pseudo-fermion
number, $\mathcal{N} = 0$, Eq.~\eqref{pf-partition0}
can be simplified as
\begin{gather}
 \label{pf-partition}
  \langle{\cal O}\rangle=\lim_{\eta\rightarrow-\infty}
  \frac{\langle{\cal O}\rangle_{\rm pf}}{\langle{\cal N}\rangle_{\rm pf}}  .
\end{gather}
The dissipative action is to be rewritten in the Keldysh
representation. We plug representation~\eqref{pseudofermion} into
the hamiltonian \eqref{ham4} and integrate out electrons in the lead
and the island. This leads to the following effective action
\begin{gather}
  \label{pseudo-action1}
   \begin{split}
   S&=
  \int dt\bar{\psi}\Big(i\partial_t-\frac{\sigma_z \Delta}{2}+\eta\Big)\psi \\
  &+\frac{g}{8}\int\bar{\psi}(t)\gamma_i\sigma_-\psi(t)\Pi_{ij}(t,t^\prime)
   \bar{\psi}(t^\prime)\gamma_j\sigma_+\psi(t^\prime)\,dtdt^\prime
   \end{split}
\end{gather}
Here $\sigma$\ stand for Pauli matrices,\
$\sigma_\pm=(\sigma_x\pm i\sigma_y)/2$,\ and
\begin{equation}
\gamma_1=\gamma_q = \begin{pmatrix}
0 &1 \\
1& 0
\end{pmatrix}, \qquad  \gamma_2=\gamma_c = \begin{pmatrix}
1 & 0 \\
0 & 1
\end{pmatrix}
\end{equation}
are matrices in Keldysh space. The
pseudo-fermion operators $\psi$\ are understood as vectors in the
tensor product of isospin and Keldysh space. $\Pi_{ij}$ stands for
the matrix of polarization operator~\eqref{aes1}-\eqref{aes2}. Next,
we write the Wigner transform of the quantum kinetic equation for
pseudo-fermion distribution function
\begin{gather}
  \label{kinetic-pseudo}
   -i\partial_\tau{\cal F}_{\varepsilon}(\tau)
   =\Sigma^{K}_\varepsilon(\tau)-\Sigma^{R}_\varepsilon(\tau){\cal F}_{\varepsilon}(\tau)+
   \Sigma^{A}_\varepsilon(\tau){\cal F}_{\varepsilon}(\tau).
\end{gather}
Here as before, we neglected all derivatives with respect to slow
time\ $\tau$. All functions entering Eq.~\eqref{kinetic-pseudo} are
understood as matrices acting in the isospin space. From the
appearance of Eq.~\eqref{kinetic-pseudo} we conclude that
characteristic relaxation time of pseudo-fermion distribution
function ${\cal F}_\varepsilon(\tau)$\ is $\tau_{\rm
pf}\sim1/(g\max\{\Delta,\,T\})$ and is much shorter than $\tau_E$.

It allows us to consider pseudo-fermions to be in the stationary state.  Then the l.h.s. of the
kinetic equation~\eqref{kinetic-pseudo} can be omitted and we obtain the equation
for pseudo-fermion
 distribution function:
\begin{gather}
  \label{kinetic-pseudo1}
  {\cal F}_\varepsilon(\tau)=
   \frac{\Sigma^K_\varepsilon(\tau)}{2i\Im\Sigma^R_\varepsilon(\tau)}.
\end{gather}

With the help of Eq.~\eqref{pseudo-action1} we write down equations
for the pseudo-fermion self-energies (Fig.~\ref{figure7}):
\begin{gather}
  \begin{split}
   &\Sigma_{+}(t,t^\prime)=
   \frac{ig}{8}\sum_{ij}\Pi^{ij}_{t^\prime t}\gamma_j G_{tt^{\prime},-}\gamma_i,\\
   &\Sigma_{-}(t,t^\prime)=
   \frac{ig}{8}\sum_{ij}\Pi^{ji}_{tt^\prime }\gamma_j G_{tt^{\prime},+}\gamma_i.
   \end{split}
\end{gather}
Here $G_{tt^{\prime},\sigma}$ stands for the pseudo-fermion Green
functions corresponding to the first line in
Eq.~\eqref{pseudo-action1} and  $\Sigma_{\pm}$ are matrices in the
Keldysh space. We will need the explicit expressions for their
Wigner transforms:
\begin{gather}
  \Sigma^K_{\varepsilon,\sigma}=-\frac{ig}{2}\int\frac{d\omega}{2\pi}\Im\Pi^R_\omega\Im
   G^{R}_{\varepsilon+\sigma\omega,-\sigma}\Big\{
       {\cal F}^{-\sigma}_{\varepsilon+\sigma\omega}B_{\omega}-\sigma\Big\},\notag \\
   \Im\Sigma^{R}_{\varepsilon,\sigma}=-\frac{g}{4}\int\frac{d\omega}{2\pi}\Im\Pi^R_\omega\Im G^{R}_{\varepsilon+\sigma\omega,,-\sigma}
  \Big\{B_\omega-\sigma{\cal
  F}^{-\sigma}_{\varepsilon+\sigma\omega}\Big\},\notag \\
  \Re\Sigma^{R}_{\varepsilon,\sigma}=-\frac{g}{4}\int\frac{d\omega}{2\pi}
  \Im\Pi^R_\omega\Re G^{R}_{\varepsilon+\sigma\omega,-\sigma}
  B_\omega.  \label{self-energy1}
\end{gather}
Here\ $\sigma$\ stands for\ $\pm$ and
\begin{equation}
G^{R}_{\varepsilon,\sigma}  = \Bigl ( \varepsilon +\eta - \frac{\Delta \sigma}{2}+i 0\Bigr )^{-1} .
\end{equation}
Combining Eqs.~\eqref{kinetic-pseudo1} and ~\eqref{self-energy1} we find the following equation
for the pseudo-fermion distribution function
\begin{gather}
  \begin{split}
   {\cal F}^\sigma_{\varepsilon}=\frac{B_{-\sigma(\varepsilon+\frac{\Delta\sigma}{2}+\eta)}
   {\cal F}^{-\sigma}-\sigma}
   {B_{-\sigma(\varepsilon+\frac{\Delta\sigma}{2}+\eta)}-\sigma
   {\cal F}^{-\sigma}}
  \end{split}
\end{gather}
where ${\cal F}^\sigma = {\cal
F}^\sigma_{\Delta\sigma/2-\eta}$ . Plugging $\varepsilon=\Delta\sigma/2-\eta$ we arrive at the closed
equation for $\mathcal{F}^\sigma$:
\begin{gather}
 \label{self-con1}
   {B}_{-\Delta}({\cal F}^\sigma-{\cal F}^{-\sigma})=
   ({\cal F}^\sigma{\cal F}^{-\sigma}-1)\sigma.
\end{gather}
Now we need to investigate
asymptotic properties of functions\ ${\cal F}^{\sigma}$\ when
pseudo-fermion chemical potential\ $\eta\rightarrow-\infty$.
It is natural to expect that the equilibrium result
\begin{equation}
\label{self-con2}
\lim\limits_{\eta\to -\infty} {\cal F}^\sigma =1
\end{equation}
 survives in the non-equilibrium.
As one can check this assumption satisfies Eq.~\eqref{self-con1}.

In order to solve the quantum kinetic equation~\eqref{ke-set}, we need
to compute $\Im {\cal D}^R_\omega$ in the strong coupling limit\ $g\ll1$. With the help of
Eqs.~\eqref{ham5} and \eqref{pseudofermion},  one easily finds in the zeroth order in $g$:
\begin{eqnarray}
   \Im{\cal D}^R_{\omega,\rm pf} &=& \int\frac{d\varepsilon}{2\pi}\Im G^{R,-}_{\varepsilon+\omega}
   \Im G^{R,+}_{\varepsilon}\Big ( {\cal F}^+_\varepsilon-{\cal F}^-_{\varepsilon+\omega}\Big ) \notag \\
   &=&
  \frac{\pi}{2}\delta(\omega+\Delta)\Big ( {\cal F}^+-{\cal
  F}^-\Big ) .
\label{cor1}
\end{eqnarray}
Now we express the physical correlation function through pseudo fermion one\
$\Im{\cal D}^R_{\omega,\rm pf}$.
By using the following zeroth order in $g$ result for the pseudo-fermion number
\begin{equation}
  \label{density1}
   \langle{\cal N}\rangle_{\rm pf}=\sum\limits_\sigma\int\frac{d\varepsilon}{2\pi}\Im
    G^R_{\varepsilon,\sigma}\Big ( {\cal F}^\sigma_{\varepsilon}-1\Big )
    =1- \frac{{\cal F}^++{\cal F}^-}{2}
    \end{equation}
we obtain
\begin{gather}
      \Im{\cal D}^R_{\omega}
      =-\pi\delta (\omega+\Delta) \lim\limits_{\eta\to -\infty}
      \frac{{\cal F}^-+1}{2 B_{-\Delta}+1 -{\cal F}^-} \notag \\
      =  -\frac{\pi\delta(\omega+\Delta)}{{B}_\omega}.
        \label{final1}
\end{gather}
Eq.~\eqref{final1} is the generalization of the equilibrium result for correlation
function\ $\Im{\cal D}^R_{\omega}$ (see Refs.~[\onlinecite{schoeller,burmistrov2}])\
over the non-equilibrium case.

Next, by using Eqs.~\eqref{ham5} and \eqref{pseudofermion},  we find in the zeroth
order in $g$:
\begin{eqnarray}
   {\cal D}^K_{\omega,\rm pf} &=&   -2 i \int\frac{d\varepsilon}{2\pi}\Im G^{R}_{\varepsilon+\omega,-}
    \Im G^{R}_{\varepsilon,+}\big(1-{\cal F}^-_{\varepsilon+\omega}{\cal F}^+_{\varepsilon}\big)
   \notag \\
   &=& \pi i \delta(\omega+\Delta) (\mathcal{F}^+\mathcal{F}^--1)
\label{cor1K}
\end{eqnarray}
Expressing the physical correlation function through pseudo-fermion
one\ ${\cal D}^K_{\omega,\rm pf}$, we obtain
\begin{gather}
  \label{final1K}
      {\cal D}^K_{\omega}
      =  -2 \pi i \delta(\omega+\Delta) .
\end{gather}
This result implies that the boson distribution function
$\mathcal{B}_\omega$ is determined by $B_\omega$ in the same as in the weak coupling regime,
\begin{equation}
\mathcal{B}_\omega = B_\omega  \label{Bto} .
\end{equation}

Before proceeding with the solution of the quantum kinetic equation we prefer to
perform one-loop renormalization of the theory. This is done to sum
up all large logarithmic corrections (which otherwise arise in perturbative
analysis) and absorb them into renormalized physical constants of
the theory.

\subsection{One-loop structure of the pseudo-fermion theory}

In this section, we establish the out-of-equilibrium generalization of
the scaling of fundamental parameters in the pseudo-fermion theory (the gap\
$\Delta$, the coupling constant\ $g$), the Green's function
 and the average pseudo-fermion density\
$\langle\mathcal{N}\rangle_{\rm pf}$. We expect  that the
action~\eqref{pseudo-action1} can be renormalized with only one
scaling parameter\ $Z$ like in the equilibrium case\
[\onlinecite{larkin,Si,Demler}]. This is indeed the case and the
obtained renormalized structure of the theory is a natural
generalization of the equilibrium one. The renormalized
pseudo-fermion Green's function becomes
\begin{gather}
  \label{green1_0}
   \overline{G}_{\varepsilon,\sigma}^{R,A}=\frac{Z(\lambda)}{\varepsilon-\bar{\xi}_\sigma\pm i\bar{g}
   \Gamma_\sigma(\varepsilon)}, \qquad
   \bar{\xi}_\sigma=-\eta+\sigma\bar{\Delta}/2 ,
\end{gather}
where
\begin{gather}
    \label{green1_1}
   Z(\lambda)=\Bigl (1+\frac{g}{2\pi^2}\lambda \Bigr )^{-1/2}, \quad
   \lambda=\int
   \frac{{B}_{\omega}}{2\omega}d\omega ,
\end{gather}
See Appendix~\ref{APPENDIX-RENORM} for details of the computation.
It is straightforward to check that coupling constant\ $g$\ and gap
$\Delta$ are renormalized according to:
\begin{gather}
  \label{green1_2}
\bar{g}=g Z^2(\lambda),\qquad \bar{\Delta}=\Delta Z^2(\lambda).
\end{gather}
To complete the renormalization picture we need to establish the
scaling dimension of the pseudo-fermion number\ $\langle
\mathcal{N}\rangle_{pf}$. In complete analogy with\
Ref.~[\onlinecite{larkin}], \ $\langle \mathcal{N}\rangle_{pf}$
happens to have  no renormalization
\begin{gather}\label{density-scaling}
 \overline{\langle{N}\rangle}_{pf} = \langle{N}\rangle_{pf}
\end{gather}
For completeness we present the rigorous proof  of
Eq.~\eqref{density-scaling} via Callan-Symanzik equation in Appendix
~\ref{APPENDIX-RENORM}.

The integral in Eq.~\eqref{green1_1} runs over frequencies $E_c\gg
|\omega|\gg \omega_0 = \max\{T_r,\varepsilon_d,\bar{\Delta}\}$. The
energy scale $\omega_0$ determines the natural scale at which the RG
procedure has to be stopped. The Green's function \eqref{green1_0}
acquires the width
\begin{gather}
  \label{width}
  \Gamma_\sigma(\varepsilon)=
   \frac{1}{8\pi}(\varepsilon-\bar\xi_{-\sigma})
   [\bar{{\cal F}}^{-\sigma}+B_{\varepsilon-\bar{\xi}_{-\sigma}}] ,
\end{gather}
where $ \bar{{\cal F}}^{\sigma}\equiv{\cal
F}^{\sigma}_{\bar{\xi}_\sigma} $.  Therefore, the renormalized
physical correlation function becomes
\begin{gather}
\Im  \mathcal{D}^R_\omega =-Z^2(\lambda)\frac{\pi\delta(\omega+\bar{\Delta})}{{B}_\omega}
\label{ImDR_RG} , \\
\mathcal{D}^K_\omega  =-2 \pi i Z^2(\lambda)\delta(\omega+\bar{\Delta})
\notag
\end{gather}

\subsection{Electron distribution relaxation in the island\label{EDrel}}

In this section we consider the relaxation in the non-equilibrium case, $\tau_E \ll \tau_{ee}$.
We focus on the most interesting case of the Coulomb peak:\ $\Delta=0$. Then the quantum kinetic
equation~\eqref{ke-set} is
greatly simplified (cf. Eq.~\eqref{KEsim1}):
\begin{gather}
   \label{ke-set1}
    \partial_\tau F_\varepsilon^d=-\frac{\mathcal{G}(\tau)\delta}{2\pi}\Big(F^d_\varepsilon-F_\varepsilon^r\Big), \\
    \mathcal{G}(\tau) = \frac{g}{2} \Bigl [ 1+\frac{g \lambda}{2\pi^2}\Bigr ]^{-1} . \label{ke-Gt}
\end{gather}
Here, we stress that the kinetic equation
\eqref{ke-set1}-\eqref{ke-Gt} is of the {\it infinite} order in the
electron distribution function on the island. Indeed, $\lambda$
involves $F_\varepsilon^d$ via electron-hole distribution function
$B_\omega$.

The formal solution reads
\begin{gather}
   F^d_\varepsilon(\tau)=F_\varepsilon^r+(F^d_\varepsilon(0)-F_\varepsilon^r)\exp\Big [-\delta\int_0^\tau
   \frac{d\tau^\prime}{2\pi} \mathcal{G}(\tau^\prime)\Big ] .
\end{gather}
The function $\mathcal{G}(\tau)$ obeys the differential equation
\begin{gather}
   \label{dif-x}
     \partial_\tau \mathcal{G}(\tau)=-\frac{\delta}{2\pi} \mathcal{G}^2(\tau)
     \Big[ \frac{ \mathcal{G}(\tau)}{ \mathcal{G}_r}-1\Big ].
\end{gather}
The solution of Eq.~\eqref{dif-x} is given as
\begin{gather}
   \label{sol-x}
\frac{\delta  \mathcal{G}_r \tau}{2\pi} = f\left (\frac{\mathcal{G}_r}{\mathcal{G}(0)} \right ) -
f\left (\frac{\mathcal{G}_r}{\mathcal{G}(\tau)} \right ), \\
f(z) = z+\ln(1-z) . \notag
\end{gather}
By using the relation
\begin{gather}
\delta\int_0^\tau
   \frac{d\tau^\prime}{2\pi} \mathcal{G}(\tau^\prime) = \frac{\delta \mathcal{G}_r\tau}{2\pi} -
   \frac{\mathcal{G}_r}{\mathcal{G}(0)}+ \frac{\mathcal{G}_r}{\mathcal{G}(\tau)}
\end{gather}
which follows from Eq.~\eqref{dif-x}, we obtain
\begin{gather}
   F^d_\varepsilon(\tau)=F_\varepsilon^r+(F^d_\varepsilon(0)-F_\varepsilon^r) \exp\Bigl [
-   \frac{\delta \mathcal{G}_r\tau}{2\pi} + \frac{\mathcal{G}_r}{\mathcal{G}(0)} -
\frac{\mathcal{G}_r}{\mathcal{G}(\tau)}
\Bigr ] .
\end{gather}

Since Eq.~\eqref{sol-x} can not be solved analytically with respect to $\mathcal{G}(\tau)$ it is
instructive to investigate limiting cases.

Let us assume that the effective energy of electrons in the island $\varepsilon_d \gg T_r$ such
that $\mathcal{G}(0) \gg \mathcal{G}_r$. Then, expanding $f(z)$ in the series in $z$, we
find
\begin{gather}
     F^d_\varepsilon(\tau)=F^r_\varepsilon+\Bigl (F^d_\varepsilon(0)-F^r_\varepsilon\Bigr )\hspace{3.5cm}{}\notag \\
     \times \exp\left [ \frac{\mathcal{G}_r}{\mathcal{G}(0)}-\sqrt{\frac{\mathcal{G}^2_r}
     {\mathcal{G}^2(0)}+\frac{\delta \mathcal{G}_r \tau}{\pi}}\right ] . \label{Fd-1}
\end{gather}
Eq.~\eqref{Fd-1} is valid provided $\mathcal{G}(\tau) \gg
\mathcal{G}_r$, i.e., for not too long times: $\tau \ll \pi /\delta
\mathcal{G}_r$. It is worthwhile to mention that standard
exponential relaxation
\begin{equation}
  F^d_\varepsilon(\tau)=F^r_\varepsilon+\Bigl (F^d_\varepsilon(0)-F^r_\varepsilon\Bigr ) \exp
  \left ( - \frac{\delta \mathcal{G}(0)\tau}{2\pi}\right ) \label{Fd-11}
\end{equation}
occurring at short time $\tau \ll \pi \mathcal{G}_r/ (\delta
\mathcal{G}^2(0))$ transforms
 into regime of slower relaxation at intermediate time $\pi
 \mathcal{G}_r/ (\delta \mathcal{G}^2(0))\ll \tau \ll \pi/(\delta \mathcal{G}_r)$:
\begin{equation}
  F^d_\varepsilon(\tau)=F^r_\varepsilon+\Bigl (F^d_\varepsilon(0)-F^r_\varepsilon\Bigr ) \exp\left ( - \sqrt\frac{\delta
  \mathcal{G}_r\tau}{\pi}\right )  .
  \label{Fd-12}
\end{equation}
At longer time $\tau \gg \pi/(\delta \mathcal{G}_r)$, function $\mathcal{G}(\tau)$ becomes
almost equal to $\mathcal{G}_r$ and we find again the regime of standard exponential relaxation:
\begin{gather}
   F^d_\varepsilon(\tau)=F_\varepsilon^r+(F^d_\varepsilon(0)-F_\varepsilon^r) \exp\Bigl [
-   \frac{\delta \mathcal{G}_r\tau}{2\pi} \label{Fd-13}
\Bigr ] .
\end{gather}

The same exponential relaxation as given by Eq.~\eqref{Fd-13} holds if the effective energy
of electrons in the island $\varepsilon_d$ is slightly larger than $T_r$ such that
$\mathcal{G}(0)-\mathcal{G}_r \ll \mathcal{G}(0), \mathcal{G}_r$.

\subsection{Temperature relaxation in the island \label{Trel}}

Now we investigate the relaxation in the quasi-equilibrium case, $\tau_E \gg \tau_{ee}$.

\subsubsection{Coulomb peak, $\Delta=0$}

We start from the regime of the Coulomb peak:\ $\Delta=0$.
In the quasi-equilibrium regime, one needs to add to the r.h.s. of Eq.~\eqref{ke-set1} the
collision integral $I^{(ee)}_{\varepsilon}$ due  to electron-electron interaction in the island.
It is this term that makes the electron distribution to be a Fermi-function. By using the well-known
identity $\int d\varepsilon \varepsilon I^{(ee)}_{\varepsilon} = 0$, we obtain the following equation:
\begin{gather}
   \label{ke-set1_2}
    \frac{d T^2_d}{d \tau}=-\frac{\mathcal{G}(\tau)\delta}{2\pi} \Bigl (T_d^2(\tau)-T_r^2 \Bigr )
\end{gather}
where $\mathcal{G}(\tau)$ is given by Eq.~\eqref{ke-Gt}. In the quasi-equilibrium case, we can
not derive closed equation for $\mathcal{G}(\tau)$ as it was done in the non-equilibrium case
due to the presence of additional term  $I^{(ee)}_{\varepsilon}$ in the r.h.s. of Eq.~\eqref{ke-set1}.

Assuming that $T_d(0) \gg T_r$ we can estimate  $\lambda$
with logarithmic accuracy as $\lambda=\ln{E_c}/{T_d}$.
Then, we find from Eq.~\eqref{ke-set1_2}
\begin{equation}
\mathcal{G}(\tau) = \mathcal{G}(0)
\Biggl [ 1+\frac{\delta\mathcal{G}^2(0) \tau }{2\pi^3} \Biggr ]^{-1/2}  \label{Gr1}
\end{equation}
and
\begin{gather}
   T_d(\tau) = T_d(0) \exp \Biggl [ \frac{\pi^2}{\mathcal{G}(0)}
    \left (1-\sqrt{1+\frac{\delta \mathcal{G}^2(0)\tau}{2\pi^3}}\right )
    \Biggr ] . \label{Gr2}
\end{gather}
The solution~\eqref{Gr2} is valid provided  the condition $T_d(\tau)\gg T_r$ holds.
If $T_d(0) \gg T_r \exp(\pi^2/\mathcal{G}(0))$,  then the exponential relaxation
\begin{equation}
 T_d(\tau) = T_d(0) \exp \Biggl [ -\frac{\delta \mathcal{G}(0)\tau}{4\pi}
    \Biggr ] \label{Gr3}
\end{equation}
developing  during initial period $\tau \ll 2\pi^3/[\delta \mathcal{G}^2(0)]$  transforms
into regime of slower relaxation at intermediate time:
\begin{gather}
   T_d(\tau) = T_d(0) \exp \Biggl [ -\sqrt{\frac{\pi \delta\tau}{2}}
    \Biggr ], \label{Gr4}
 \\
      \frac{2\pi^3}{\delta \mathcal{G}^2(0)}\ll \tau \ll \frac{2}{\pi \delta} \ln^2
      \frac{T_d(0)}{T_r} .
    \notag
    \end{gather}
We mention that in this regime the temperature relaxation is
independent of the quantity $\mathcal{G}(0)$ which determines the
SET conductance. In the opposite case, $T_d(0) \ll T_r
\exp(\pi^2/\mathcal{G}(0))$ the temperature $T_d(\tau)$ evolves
according to Eq.~\eqref{Gr3} for $\tau \ll (4\pi / \delta
\mathcal{G}(0)) \ln T_d(0)/T_r$ .

\begin{figure}[t]
  \includegraphics[width=80mm]{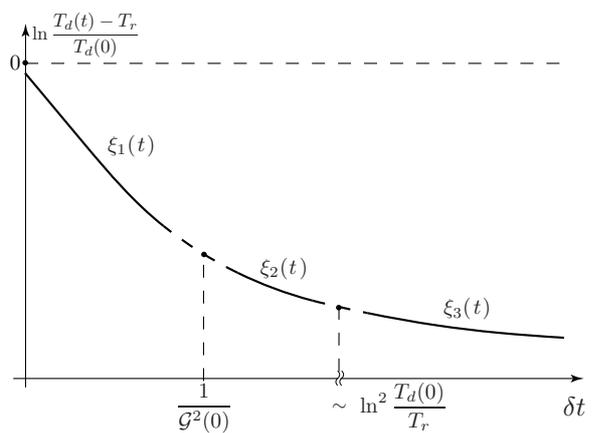}
  \caption{\label{figure4}
    The dynamics of temperature relaxation, $g\ll1$,\
    $T_d(0)\gg T_r\exp\big(\pi^2/{\cal G}(0)\big)$. Here, $\xi_1(t)\sim-{\cal G}(0)\delta t,\ \
    \xi_2(t)\sim-\sqrt{\delta t},\ \ \xi_3(t)\sim-{\cal G}_r\delta t.$
          }
\end{figure}

 At longer times  $\tau \gg (4\pi / \delta \mathcal{G}_r) \ln T_d(0)/T_r$ the
 temperature $T_d(\tau)$ becomes of the order of $T_r$: $T_d(\tau)-T_r \ll T_r$ and we find the
 standard exponential relaxation:
\begin{equation}
 T_d(\tau) = T_r +(T_d(0)-T_r)  \exp \Biggl ( -\frac{\delta \mathcal{G}_r\tau}{4\pi}
    \Biggr ) . \label{Gr5}
\end{equation}
Evolution of the temperature of electrons in the island is presented
in Fig.~\ref{figure4}. We mention universality of the relaxation at
long time when the difference between the electron distribution in
the island and in the reservoirs becomes small. In non-equilibrium
$\tau_E \ll \tau_{ee}$ and quasi-equilibrium $\tau_E \gg \tau_{ee}$
regimes the relaxation is exponential with a rate of the order of
$\delta \mathcal{G}_r$. The same exponential relaxation as in
Eq.~\eqref{Gr5} holds if the temperature of electrons in the island
$T_d(0)$ is slightly larger than $T_r$, $T_d(0)-T_r \ll T_d(0),
T_r$.

It is worthwhile to mention that there is a parametric region of
time domain\ $\mathcal{G}_r/(\delta{\cal
G}(0)^2)\ll\tau\ll1/(\delta{\cal G}(0)^2)$, when the relaxation of
the distribution function in the non-equilibrium regime is much
slower
$\ln(F^d_\varepsilon(\tau)-F^r_\varepsilon)/(F^d_\varepsilon(0)-F^r_\varepsilon)\sim-\sqrt{\tau}$
than the relaxation of the (Fermi) distribution function in the
quasi-equilibrium regime, i.e., relaxation of temperature,
$\ln(T_d/T_d(0))\sim-\tau$ .

\subsubsection{Coulomb valley, $\bar\Delta \gg T_d(0)$}

Now we consider the relaxation of the electron temperature on the island in the regime of Coulomb
valley, $\bar\Delta \gg T_d(0)$. By using Eq.~\eqref{ImDR_RG}, we rewrite the quantum kinetic
equation~\eqref{ke-set} as
\begin{gather}
   \label{ke-set1_tr}
    \partial_\tau F_\varepsilon^d=\frac{\bar g \delta}{4\pi}\Biggl ( F_{\varepsilon+\bar\Delta}^r-F^d_\varepsilon +
    \frac{1 - F_\varepsilon^d F_{\varepsilon+\bar\Delta}^r}{B_{-\bar\Delta}}\Biggr ) .
 \end{gather}
We remind that we consider the quasi-equilibrium regime.
Then we need to add to the r.h.s. of Eq.~\eqref{ke-set1_tr}
the collision integral $I^{(ee)}_{\varepsilon}$ which describes scattering
due to electron-electron interaction in the island. In what follows we
assume that the condition $T_d(0) \gg T_r$ holds.
With the help of the following results
\begin{gather}
   \int d\varepsilon \,\varepsilon (1-F^d_\varepsilon F^r_{\varepsilon-\omega})
   =T_d^2  \sgn\omega\bigg[
   \frac{\omega^2}{T_d^2}-\frac{\pi^2}{3} \hspace{2cm}{}\notag \\
  -4{\rm li}_2(-e^{-|\omega|/T_d})  +\frac{4|\omega|}{T_d}\ln\Big(1+e^{-|\omega|/T_d}\Big)
   \bigg] ,  \label{int1} \\
  \label{int2}
  B_\omega=\frac{2T_d}{\omega} \ln \left ( 2\cosh\frac{\omega}{2T_d} \right )
\end{gather}
which are valid for $T_r\ll T_d$ (${\rm li}_2(z) = \sum_{n=1}^\infty z^n/n^2$ denotes the
polylogarithmic function), we obtain from Eq.~\eqref{ke-set1_tr}
\begin{gather}
  \label{ke-set2}
    \frac{d T_d}{d \tau} =-\frac{3\delta \mathcal{G}(\tau)}{4\pi^3} T_d(\tau), \\\
    \mathcal{G}(\tau) = \frac{\bar{g} \bar{\Delta}}{T_d(\tau)}
    \exp \left (-\frac{\bar\Delta}{T_d(\tau)}\right )  .\label{Gwr}
\end{gather}
We can estimate parameter $\lambda$ with logarithmic accuracy and find  $\lambda =
\ln E_c/\bar\Delta$ since the temperature of electrons in the island $T_d \ll \bar{\Delta}$.
Therefore, both $\bar{g}$ and $\bar{\Delta}$ are independent of $\tau$. Integration of
Eq.~\eqref{ke-set1_tr}  yields
\begin{gather}
\frac{3\delta \bar{g} \tau}{4\pi^3} = h\left (\frac{\bar\Delta}{T_d(0)}\right )-h
\left (\frac{\bar\Delta}{T_d(\tau)}\right ) \label{Gd1}\\
h(z) =  e^z/z - {\rm Ei}(z) .
\end{gather}
Here ${\rm Ei}(z) = - \int_{-z}^\infty dt \exp(-t)/t$ stands for the
integral exponential. By using the asymptotic $h(z) = \exp(z)/z^2$
at $z \gg 1$, we obtain
\begin{gather}
      \mathcal{G}(\tau)=\mathcal{G}(0)
      \Biggl [1+ \frac{T_d(0)}{\bar{\Delta}} \ln\Big(
       1+ \frac{3\delta \bar\Delta \mathcal{G}(0) \tau}{4\pi^3 T_d(0)}
      \Big) \Biggr ]^{-1}
      \notag \\
 \times  \Biggl [1+ \frac{3\delta \bar\Delta \mathcal{G}(0) \tau}{4\pi^3 T_d(0)}
 \Biggr ]^{-1} \label{Gt-CP}
\end{gather}
and
\begin{gather}
      T_d(\tau)=T_d(0)
      \Biggl [1+ \frac{T_d(0)}{\bar{\Delta}} \ln\Big(
       1+ \frac{3\delta \bar\Delta \mathcal{G}(0) \tau}{4\pi^3 T_d(0)}
      \Big) \Biggr ]^{-1} . \label{Td-CP}
\end{gather}
The results~\eqref{Gt-CP} and \eqref{Td-CP} are valid at not too long times
\begin{equation}
\tau \ll \frac{4\pi^3 T_d(0)}{3\delta \bar\Delta \mathcal{G}(0)} \exp
\left (\frac{\bar\Delta}{T_r}\right ) .
\end{equation}
 As expected, due to the exponentially small SET conductance in the
sequential tunneling regime, the temperature relaxation is very
slow, namely, logarithmical. Therefore, it is instructive to
consider contribution to the temperature relaxation due to the
electron co-tunneling.

\subsubsection{Inelastic cotunneling regime}

As known very well, due to exponential suppression of the sequential
tunneling mechanism deep in the Coulomb valley, $T_d \ll
\bar{\Delta}$, the higher order process of inelastic cotunneling
dominates the transport.~\cite{nazarov} Contrary to the case of
sequential tunneling, the cotunneling contribution to the collision
integral in the r.h.s. of Eq.~\eqref{ke-set} comes from frequencies
of order $\omega\sim T_d\ll\bar{\Delta}$.

In the pseudo-fermion technique the inelastic cotunneling
is revealed as the broadening of
delta-peaks in the imaginary part of the retarded and advanced
pseudo-fermion Green functions.~\cite{schoeller,burmistrov2}
After taking into account Eq.~\eqref{width}, the integrand in Eq.~\eqref{cor1}
becomes of a complex pole structure. There are two pairs of proximal poles
\begin{gather}
   \begin{split}
   \varepsilon&=\xi_+\pm i\bar{g}\Gamma_+(\bar{\xi}_+), \\
   \varepsilon&=\xi_--\omega\pm i\bar{g}\Gamma_-(\bar{\xi}_-) .
   \end{split}
\end{gather}
There is an
additional series of Matsubara-type poles resulting from
distribution functions $\mathcal{F}_\epsilon^+$ and $\mathcal{F}_{\epsilon+\omega}^-$.
They lead to logarithmically divergent sums. The
latter are controlled by the renormalization scheme. In our case all
leading logarithms are absent. They have already been absorbed into
renormalized constants $\bar{g}$ and $\bar{\Delta}$\ by the proper
choice of reference energy scale. Thus we can omit all divergent
sums over Matsubara frequencies. Expanding in the $\omega/\bar \Delta$ we
obtain
\begin{gather}
   \Im{\cal D}^R_{\omega,\rm pf}=
   \frac{\bar g Z^2\omega}{8\pi}\frac{{\cal F}^++{\cal
   F}^-}{\bar \Delta^2},\ \ |\omega|\ll|\bar\Delta| .
\end{gather}
Next we use the same arguments that led us to leading order
espression~\eqref{final1}. The function $\Im{\cal D}^R_\omega$\ then
reads
\begin{gather}
  \label{cot1}
    \Im{\cal D}^R_\omega=-\frac{\bar g Z^2}{4\pi}\frac{\omega}{\bar \Delta^2},\ \
   |\omega|\ll|\bar \Delta| .
\end{gather}
Using Eq.~\eqref{cot1}, we rewrite the quantum kinetic equation~\eqref{ke-set} as
\begin{gather}
   \label{ke-set1_tr_co}
    \partial_\tau F_\varepsilon^d=\frac{\bar g^2 \delta}{16\pi^3 \bar {\Delta}^2}\int d\omega\,
    \omega \Biggl [ \left ( F_{\varepsilon-\omega}^r-F^d_\varepsilon\right ) B_\omega + 1 - F_\varepsilon^d
    F_{\varepsilon-\omega}^r\Biggr ].
 \end{gather}
We remind that we consider the quasi-equilibrium regime. We mention
that Eq.~\eqref{ke-set1_tr_co} coincides with the kinetic equation
derived for the cotunneling regime in
Ref.~[\onlinecite{BaskoKravtsov}]. Then we need to add to the r.h.s.
of Eq.~\eqref{ke-set1_tr_co} the collision integral $I^{(ee)}_{\varepsilon}$
which describes scattering due to electron-electron interaction in
the island.

 In the case of $T_d -T_r\ll T_r$, we obtain
\begin{equation} \label{Eqcot1}
\frac{d T_d}{d\tau}  = - \frac{3 \delta \mathcal{G}_r}{5\pi}  (T_d-T_r)
\end{equation}
where $\mathcal{G}_r = \bar{g}^2 T_r^2/(6 \bar\Delta^2)$ stands for the
equilibirum SET conductance in the cotunneling approximation.
In the opposite case $T_d\gg T_r$, by using Eq.~\eqref{ke-set1_tr_co},
we find the following equations:
\begin{gather}
 \frac{d T_d}{d \tau} =
 - \frac{3 \delta \mathcal{G}(\tau)}{20\pi} T_d(\tau), \label{Eqco1}\\
  \mathcal{G}(\tau) = \frac{\bar{g}^2T_d^2(\tau)}{6 \bar{\Delta}^2}  \label{Eqco2}
\end{gather}
which govern the temperature relaxation.
It is worthwhile to mention that if one substitutes $\mathcal{G}_r$ by $\mathcal{G}(\tau)$
in Eq.~\eqref{Eqcot1} then it becomes similar to Eq.(4) of Ref.~[\onlinecite{beloborodov4}]
 for $V=0$ and in the absence of phonons. However, due to different numerical coefficients
 in the right hand side of Eqs.~\eqref{Eqcot1} and \eqref{Eqco1} such substitution is
 impossible even on the level of interpolating expression. Therefore, in the case
 $T_d-T_r\sim T_r$ one needs to solve Eq.~\eqref{ke-set1_tr_co} numerically.

Though, Eq.~\eqref{Eqcot1} leads to the standard exponential relaxation, Eq.~\eqref{Eqco1} results in the relaxation according to the power-law. Indeed,
 solving Eqs.~\eqref{Eqco1}-\eqref{Eqco2}, we obtain
\begin{equation}
\mathcal{G}(\tau) = \mathcal{G}(0) \left (1+\frac{3 \delta \mathcal{G}(0)\tau}{10\pi} \right )^{-1}
\label{Eqco3}
\end{equation}
and
\begin{equation}
T_d(\tau) = T_d(0) \left [1+\frac{3 \delta \mathcal{G}(0)\tau}{10\pi} \right ]^{-1/2} \label{Eqco4}
\end{equation}
Eqs.~\eqref{Eqco3} and \eqref{Eqco4} are valid at times
\begin{equation}
\tau\ll \frac{10\pi}{3 \delta \mathcal{G}(0)}\frac{T_d^2(0)}{T_r^2} .
\end{equation}

\section{Discussions and conclusions\label{DISCUSSION}}

We have studied the relaxation dynamics of the SET under essentially
non-equilibrium conditions. The language of kinetic equations
happened to be the most adequate for this task. Analytical results
are procured in the limiting cases of weak $g\gg1$ and strong
$g\ll1$ Coulomb blockade. All relaxation equations (see
Eqs.~\eqref{KEsim1},\eqref{ke-set1},\eqref{ke-set1_2},\eqref{ke-set2})
obtained in the course reveal a pleasant generality. Namely,
\begin{gather}
\label{diss1}
     \dot{X}_d\sim- \delta \mathcal{G}(X_d) (X_d-X_r).
\end{gather}
Here $X_d$ is a relaxing physical quantity (temperature,
distribution function), and $\mathcal{G}(X_d)$ is conductance of a
SET which depends on $X_d$. Equation~\eqref{diss1} has a transparent
intuitive interpretation. Namely, the characteristic time scale
determined by the r.h.s. of Eq.~\eqref{diss1} is simply a dwell time
of a particle inside the metallic island,~\cite{Bagrets} i.e.
$\tau_E^{-1}\sim \mathcal{G}\delta$.  The inverse dwell time can be
also estimated as the ratio of the thermal conductance $\kappa$ to
the heat capacitance of the island. The latter is proportional to
$T_d/\delta$. The generality of Eq.~\eqref{diss1} is, however,
deceptive as it leads to drastically different evolution of physical
quantities over time in the case of small and large values of $g$.

\begin{figure}[t]
  \includegraphics[width=60mm]{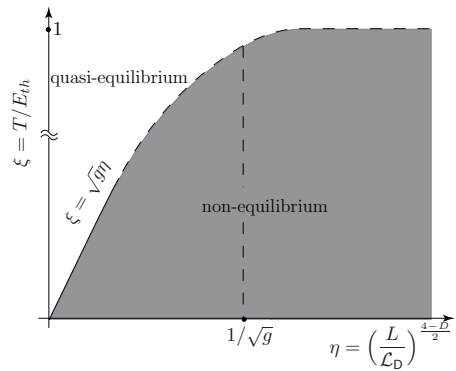}
  \caption{\label{figure5}
 Schematic diagram of different regimes for $g\gg 1$. The non(quasi)-equilibrium regime
 dominates in (un)shaded region.
          }
\end{figure}

\begin{figure}[t]
  \includegraphics[width=85mm]{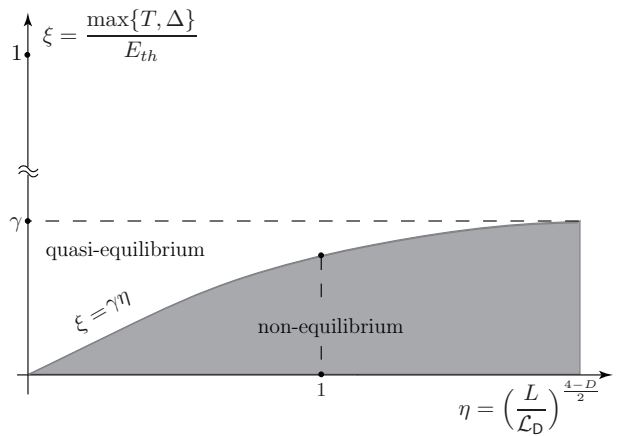}
  \caption{\label{figure6}
Schematic diagram of different regimes for $g\ll 1$. The non(quasi)-equilibrium regime dominates
in (un)shaded region. At $\Delta=0$, $\gamma=\sqrt{g}$ and $\gamma=g$ in the cotunneling case for
$T\ll\Delta$.
    }
\end{figure}

In the course of all our analysis we generally discarded the
influence of electron-phonon (e-ph) interaction. The reasoning
behind this is as follows. The e-ph scattering rate was well studied
for 2-dimensional electron gas with disorder.~\cite{mittal} The following estimate has been found
\begin{gather}
   \label{tau-e-ph}
    \tau^{-1}_{\rm e-ph}\approx8.3\times10^8\,T^3\ [s^{-1}K^{-3}].
\end{gather}
The electron-electron (e-e) scattering rate in mesoscopic systems is
widely studied as well (see, e.g.[\onlinecite{Blanter}]). For small
diffusive electron systems and for $T\ll E_{\rm Th}$ there are two
parametrically different situations~\cite{SivanImryAronov,Blanter}
\begin{gather}
   \label{tau-fermi}
   \tau^{-1}_{\rm ee}\sim
   \frac{T^2\delta}{E_{\rm Th}^2},\ \
   L\gg \mathcal{L}_{\mathbb{D}},\ \ \\
   \label{tau-diffuse}
   \tau^{-1}_{\rm ee}\sim \frac{T^2}{E_F},\ \
   L\ll \mathcal{L}_{\mathbb{D}},
\end{gather}
where $ {\cal L}_{\mathbb{D}}=(k_F l)^{\frac{2}{4-\mathbb{D}}}/k_F$
and $L$ stands for the size of the island.
Equation~\eqref{tau-diffuse} is a typical Fermi-liquid expression
coming from large
momenta of the order of the  inverse screening length.
The
upper one comes from momenta\ $k\sim1/L$ and of diffusive origin.
Let us address the question which kind of dissipation dominates in
various parametric regimes. The relaxation due to electron tunneling
can be roughly estimated as
\begin{gather}
 \label{relax-E}
  1/\tau_E \sim {\cal G} \delta.
\end{gather}
By comparing Eqs.~\eqref{tau-fermi}, \eqref{tau-diffuse} and
\eqref{relax-E}, one can see that both quasi-equilibrium and
non-equilibrium regimes can occur for $g\gg 1$ and $g\ll 1$. The
non-equilibrium regime prevails for $g\gg 1$ while the
quasi-equilibrium one dominates for $g\ll 1$ (see
Figs.~\ref{figure5} and~\ref{figure6}).

To estimate e-e scattering rate we use some experimental data taken from
the experiment by Pasquer {\it et al}.~\cite{glattli} where they
studied the Coulomb blockade effects in a small island of two-dimensional electron gas.
The experimental data were as follows: the level spacing $\delta\approx85\ mK$,
the Fermi energy $E_F\approx47\ K$, the elastic mean free path\
$l\approx15\ \mu m$, the size of an island\ $L\approx 1\ \mu
m$. This allows us to estimate the Thouless energy\ $E_{\rm Th}\approx8\
K$ and e-e relaxation rate
\begin{gather}
    \label{tau-ee-1}
   \tau_{\rm ee}^{-1}\approx1.7\times10^{8}\,T^2\ [K^{-2}s^{-1}].
\end{gather}
The typical temperature of the contemporary mesoscopic experiment
is\ $T\lesssim 100\ mK$. As we see, with lowering temperature the
e-ph scattering rate decays faster then the corresponding
electron-electron (e-e) rate. On the other hand for the same
metallic island the typical relaxation rate due to electron escape
to reservoirs is
\begin{gather}
    \label{tau-dw}
    \tau_{E}^{-1}\sim g\times10^8\ [s^{-1}]
\end{gather}

Estimates~\eqref{tau-e-ph}-\eqref{tau-dw} show that for all relevant
experimental temperatures the phonons are frozen and e-ph
interaction can safely be discarded. Next, comparing
estimates~\eqref{tau-ee-1} and~\eqref{tau-dw} we conclude that
varying $g$ two different parametric regimes explored in this paper
can indeed be realized in the experiment. Namely, fully
non-equilibrium regime is realized when $g$ is large enough and\
electron distribution function is arbitrary inside the island. The
quasi-equilibrium regime persists in the opposite limit, when $g$ is
small enough.

 In addition to the relaxation of the electron distribution in the
 island due to escape of electrons to the reservoirs
 which we consider in details above there is another mechanism of
 energy relaxation which is due to interaction $U_{ir}$ of electrons in the
 island with electrons in the reservoirs. For a sake of simplicity
 we assume that the typical interaction parameter $r_s \sim 1/(k_F a_B) \sim 1$
 with $a_B$ standing for Bohr radius. In the case $L \ll \mathcal{L}_\mathbb{D}$, the energy
 relaxation rate due to interaction $U_{ir}$ of electrons in the island with electrons in the
 reservoirs can be estimated as
\begin{equation}
\frac{1}{\tau_{ee}^{(ir)}} \sim  [\nu_d U_{ir}(k_F)]^2 \frac{T^2}{E_F} .
\end{equation}
Here $U_{ir}(k)$ denotes the Fourier transform of the interaction $U_{ir}(\bm{r})$. Provided
the condition $k_F d\gg 1$ holds $U_{ir}(k_F)$ is strongly suppressed, $\nu_d U_{ir}(k_F) \ll 1$,
and
\begin{equation}
\frac{1}{\tau_{ee}^{(ir)}} \ll \frac{1}{\tau_{ee}}.
\end{equation}
In the opposite case of large island, $L\gg \mathcal{L}_\mathbb{D}$, and for $a_B\ll L$ the
estimate for the energy relaxation rate $1/\tau_{ee}^{(ir)}$ becomes
\begin{gather}
  \label{relax-rate1}
  \frac{1}{\tau_{ee}^{(ir)}}\sim\frac{U^2_{ir}(k_L)}{U^2(k_L)-U^2_{ir}(k_L)}\frac{1}{\nu_d U(k_L)}
  \frac{
  T^2}{E_{Th}^2} \delta.
\end{gather}
Here $k_L \sim 1/L$ and $U(k)$ stands for the Fourier
transform of interaction between electrons in the island $k=0$ component of which leads to the
charging term $H_c$ in the hamiltonian \eqref{ham1}. As one can see, both cases of
$\tau_{ee} \gg
\tau_{ee}^{(ir)}$ and $\tau_{ee} \ll   \tau_{ee}^{(ir)}$ are
possible for $L\gg \mathcal{L}_\mathbb{D}$.

For $d\ll L$ where $d$ stands for the typical size of the tunneling junction between the island
and reservoir Eq.~\eqref{relax-rate1} can be simplified as
\begin{gather}
  \label{relax-rate1_1}
  \frac{1}{\tau_{ee}^{(ir)}}\sim\frac{a_B}{d} \frac{
  T^2}{E_{Th}^2} \delta.
\end{gather}
For the experiments by Pasquer {\it et al}.~\cite{glattli} we
estimate the Bohr radius $a_B\approx 10\, nm$ and  assume typical
$d$ to be of the order of $100\, nm$. Therefore, we expect that the
regime in which the main mechanism of the energy relaxation of
electrons in the island is due to its escape to the reservoirs can
be realized in a laboratory.

To summarize, we have explored heat transport and relaxation
processes in a SET with large number of tunneling channels over a
wide range of parameters. In the regime of linear response we
obtained analytical expressions for transport coefficients
(conductance, thermal conductance and the response of electric
current to temperature difference) in the entire span of values of\
$g$. It is possible to shape the general relations for linear
response coefficients into Fermi-liquid type form. There is however
an important difference, namely: the tunneling density of states
undergoes dramatic renormalization due to Coulomb interaction. The
latter leads to violation of Wiedemann-Franz law: in the\ $g\gg1$
limit the Lorentz ratio\ $\mathcal{L}$ acquires weak periodic
dependence on gate voltage (the precursor of Coulomb blockade). The
method of quantum kinetic equation supplemented with non-equilibrium
AES action has allowed us to treat Coulomb interaction exactly. We
have obtained the time evolution of electron temperature (in the
quasi-equilibrium regime) and the distribution function (in the
non-equilibrium regime) of a SET island due to particle escape to
reservoir. The corresponding collision integral is always non-local
in energy due to inelastic nature of tunneling processes: the
radiation of plasmon\ $\varphi$ always accompanies the tunneling event.
In general, this leads to highly complicated integro-differential
kinetic equations. Surprisingly we have shown that kinetic equations
can be reduced to simple differential ones in a number of wide
parametric regimes, namely: $g\gg1$ (weakly blockaded SET)\ and\
$g\ll 1$ (strongly blockaded SET\ in sequential tunneling
approximation with renormalization taken into account).
This simplification is achieved due to the
presence of strong scale separation in the problem\ $g\delta\ll T_d$ or
$g\delta\ll \varepsilon_d$. Indeed, the characteristic frequency at which
the distribution function in the kinetic equation
changes is\ $\omega\sim g\delta$, while the scale at which the
renormalization due to the presence of Coulomb interaction occurs is
$\omega\gtrsim T_d$ or $\omega\gtrsim \varepsilon_d$. This separation is that
allows us at first to treat  Coulomb interaction and secondly to study
evolution of the distribution function.

Still, quantum fluctuations of charge significantly change the
relaxation laws comparing to simple exponential ones which are
characteristic of semi-classical physics for $g\gg1$ and of orthodox
theory for\ $g\ll1$. The regime \ $g\ll1,\ \Delta\gg T$ is dominated
by cotunneling process. In the latter case the kinetic equation
retains its integro-differential structure and is to be solved
numerically elsewhere. Measurements of the predicted relaxation
dynamics are an experimental challenge.

\begin{acknowledgements}

The authors are grateful to D. Bagrets, D. Basko, I. Gornyi, A.
Ioselevich, V. Kravtsov, Yu. Makhlin, J. Pekola and K. Tikhonov for
stimulating discussions. The research was funded in part by the
Russian Ministry of Education and Science under the Contract No. P926, the Council for Grant of
the President of Russian Federation (Grant No. MK-125.2009.2), RFBR
(Grant No. 09-02-92474-MHKC and No. 07-02-00998) and RAS Programs
``Quantum Physics of Condensed Matter'' and ``Fundamentals of
nanotechnology and nanomaterials'' and also by the U.S. Department
of Energy Office of Science under the Contract No. DE-AC02-06CH11357. I.S.B. is grateful to the Low Temperature Laboratory
at Aalto University for hospitality.

\end{acknowledgements}

\appendix

\section{Keldysh form of AES action \label{APPENDIX-AES}}

For a benefit of a general reader, we outline here the details of the derivation of the Keldysh form
of AES-action from hamiltonian~\eqref{ham1}-\eqref{ham3}. To get rid
of unsuitable quartic electron-electron interaction
term~\eqref{ham3} we decouple it via Hubbard-Stratonovich bosonic
field\ $\varphi(t)$. After that the initial electron operators
are gauge-transformed according to
\begin{gather}
  d^\dag_\alpha(t)\rightarrow d^\dag_\alpha(t)
e^{i\varphi(t)},\qquad  d_\alpha(t)\rightarrow d_\alpha(t)
e^{-i\varphi(t)} ,\label{gauge-transform}
\end{gather}
the action of the system becomes gaussian in fermions:
\begin{gather}
  \begin{split}
      S&=S_0+S_c+S_t , \\
    S_0&=\int_\gamma\sum_\alpha d_\alpha^\dagger(i\partial_t-\varepsilon_\alpha^{(d)})d_\alpha\\
    &+\int_\gamma\sum_k a_k^\dagger(i\varepsilon_t-\varepsilon_k^{(r)})a_k , \\
    S_c&=\frac{1}{4E_c}\int_\gamma\dot{\varphi}^2dt +q \int_\gamma\dot{\varphi}dt , \\
    S_t&=-\int_\gamma\sum_{k,\alpha}\Big(t_{k\alpha}a_k^\dagger d_\alpha
    e^{i\varphi}+\hbox{H.c.}\Big)dt .
  \end{split}
\end{gather}
Here, for a sake of simplicity, we consider an island connected to a single reservoir.
Superscript $d$ refers to the island and\ $r$ - to the reservoir.
The integrals are understood as contour ones and\ $\gamma$ is
the Keldysh contour. Integrating out fermions we obtain the effective action for the bosonic field $\varphi$:
\begin{equation}
S_{\rm eff} = -i {\rm tr}\,\ln(\widehat{G}^{-1}+\widehat{T})+S_c .
\end{equation}
Here, matrices $\widehat{G},\ \widehat{T}$ have the following structure in the
reservoir-island space:
\begin{gather}
  \begin{split}
    &\widehat{G}=\begin{pmatrix}
                          G_{k,d}\ &\ 0\\
                          0    \ &\ G_{\alpha,r}
                       \end{pmatrix},\ \
   \widehat{T}=        \begin{pmatrix}
                          0\ &\ t_{k \alpha}X\\
                          t^\dagger_{\alpha k}X^\dagger\ &\ 0
                       \end{pmatrix},\\
                       &X=\frac{1}{\sqrt{2}}
                        \begin{pmatrix}
                          X_c\ &\ X_q\\
                          X_q\ &\ X_c
                       \end{pmatrix} ,
   \end{split}
\end{gather}
where $X_{c,q}$\ are defined in Eq.\eqref{Xs}. Expanding $S_{\rm eff}$ to the second order in $\widehat{T}$, we find
\begin{gather}
      S_{\rm eff}=\frac{i}{2}{\rm
      tr}\,\Big[\widehat{G}\widehat{T}\widehat{G}\widehat{T}\Big]+S_c .
\end{gather}
This expansion is valid in the limit $g_{ch}\ll 1$ and $N_{\rm ch}\gg1$. Computing all the traces we recover the dissipative part of
AES-action in form~\eqref{aes1} with the polarization operator $\Pi$\
given by the following general expressions:
\begin{gather}
  \label{pol1}
 \begin{split}
     \Pi^{R,A}(t,t^\prime)& = \frac{i}{2 g}\sum_{k,\alpha}|t_{k\alpha}|^2\Big(G_{k,r}^K(t^\prime,t)
     G_{\alpha,d}^{R,A}(t,t^\prime)\\
     &+G_{k,r}^{A,R}(t^\prime,t)G_{\alpha,d}^K(t,t^\prime)\Big),\\
     \Pi^K(t,t^\prime)& = \frac{i}{2 g}\sum_{k,\alpha}|t_{k\alpha}|^2\Big(G_{k,r}^K(t^\prime,t)
     G_{\alpha,d}^K(t,t^\prime)\\
     &+G_{k,r}^R(t^\prime,t)G_{\alpha,d}^A(t,t^\prime)+G_{k,r}^A(t^\prime,t)
     G_{\alpha,d}^R(t,t^\prime)\Big).
   \end{split}
\end{gather}
Provided the density of states of electrons on the island and in the reservoir are slow varying near the Fermi energy,
we can perform the summation over $\alpha$ and $k$ with the help of Eqs.~\eqref{gkk}-\eqref{conductance-def1} and
reproduce the kernel of the action in form of Eq.~\eqref{aes2}.


\section{Electron's self-energy \label{APPENDIX-2}}

Here, we present the expressions for electron's self-energy to
substantiate the derivation of the kinetic equation in
Sec.~\ref{KINETIC}. As follows from Fig.~\ref{figure2} a the lowest
order (in $1/N_{\rm ch}$) contribution to the electron's self-energy
is given by
\begin{gather}
 \begin{split}
  &\Sigma^{R,A}(t,t^\prime) = i \sum\limits_{\alpha\alpha^\prime} (2\pi)^2
  [\delta(\epsilon_\alpha)\delta(\epsilon_{\alpha^\prime})]^{1/2}\\
  & \times \sum\limits_{k} t^\dag_{\alpha k}t_{k\alpha^\prime} \Bigl [  {
    G}^{R,A}_{k,r}(t,t^\prime)
  {\cal D}^K(t,t^\prime) + {G}^{K}_{k,r}(t,t^\prime) {\cal D}^{R,A}(t,t^\prime)\Bigr ]\\
  &\Sigma^{K}(t,t^\prime) = i \sum\limits_{\alpha\alpha^\prime}
  (2\pi)^2
  [\delta(\epsilon_\alpha)\delta(\epsilon_{\alpha^\prime})]^{1/2}\\
  &\times \sum\limits_{k}
  t^\dag_{\alpha k}t_{k\alpha^\prime}\Bigl [ {G}^{K}_{k,r}(t,t^\prime)
  {\cal D}^K(t,t^\prime) + [{G}^{R}_{k,r}(t,t^\prime) -
  {G}^{A}_{k,r}(t,t^\prime)]\\
  &\times[{\cal D}^{R}(t,t^\prime)-{\cal D}^{A}(t,t^\prime)]\Bigr ]
 \end{split}
\end{gather}
In case of constant densities of states in the island and the reservoir, they can be simplified with
the help of Eqs.~\eqref{gkk}-\eqref{conductance-def1} and, then,
written in the form of Eq.~\eqref{self-energies2}.

\section{Tunneling density of states on the island \label{AppTDOS}}

 The tunneling density of states of
electrons inside the island is defined via corresponding full
retarded Green's function of original fermionic operators:
\begin{gather}
  \begin{split}
   i{\bf G}^R_d(t,t^\prime)&=
     \frac{1}{2}\big\langle d_+\bar{d}_+^\prime e^{-i(\varphi_+-\varphi_+^\prime)}
   -d_+\bar{d}_-^\prime
   e^{-i(\varphi_+-\varphi_-^\prime)}
   \\
  &+d_-\bar{d}_+^\prime e^{-i(\varphi_--\varphi_+^\prime)}-d_-\bar{d}_-^\prime
   e^{-i(\varphi_--\varphi_-^\prime)}\big\rangle\\
    &=-\frac{1}{2}\big\{
G^R_{t,t^\prime}D^K_{t^\prime,t}+G^K_{t,t^\prime}D^A_{t^\prime,t}
    \big\}
  \end{split}
\end{gather}
 Here, operators $d_{\pm}\equiv d_{\pm}(t),\ d^\prime_{\pm}\equiv
d_{\pm}(t^\prime)$ are the gauge transformed operators of electrons
inside the island (see Eq.~\ref{gauge-transform}.) Subscripts\ $\pm$
correspond to upper(lower) branch of Keldysh contour.

Switching to Wigner transform we obtain
\begin{gather}
    {\bf G}^R_d(\varepsilon)={-}\sum_\alpha \int \Big\{G^R_{\alpha,d}(\varepsilon+\omega) \mathcal{B}_\omega\Im \mathcal{D}^R_\omega
    \notag \\
     +\mathcal{D}^A_\omega F^d_{\varepsilon+\omega}
    \Im G^R_d(\varepsilon+\omega)\Big\}\frac{d\omega}{2\pi}.
\end{gather}
Then, the tunneling density of states of electrons on the island becomes
\begin{gather}
  \label{dos1}
   \nu_d(\varepsilon)=-\frac{1}{\pi}\Im {\bf G}^R_d(\varepsilon)=\nu_d\int\Im
   \mathcal{D}^R_\omega\Big\{\mathcal{B}_\omega-F^d_{\varepsilon+\omega}\Big\}\frac{d\omega}{2\pi}.
\end{gather}
Eq.~\eqref{dos1} gives the tunneling density of states of electrons on the island in a non-equilibrium regime with arbitrary electron distribution function $F_d$. In the equilibrium, it leads to the result~\eqref{dos}.


\section{Renormalization of AES-action at $g\gg 1$.\label{RENORM}}

In this appendix we present details of derivation of Eq.~\eqref{rg1} which describes renormalization of $g$ under non-equilibrium
conditions in the weak-coupling regime. According to general philosophy behind renormalization we
successively integrate partition-function over high-energy
components of field $\varphi$. We split the scalar field into slow
and fast components\ $\varphi\rightarrow\varphi+\theta$, where
$\varphi=(\varphi_c,\varphi_q)$ and $\theta=(\theta_c,\theta_q)$,
and expand the action up to quadratic order in the fast field\ $\theta$:
\begin{gather}
  \label{action1}
   \begin{split}
     &S[\varphi]\rightarrow S[\varphi]+\int b_t[\varphi] \theta(t)\,dt+\frac{1}{2}\int
     \theta(t)K^{-1}_{t,t^\prime}[\varphi]\theta(t^\prime)\,dtdt^\prime,\\
     &b_t[\varphi]=\frac{\delta S}{\delta\varphi(t)}\Big|_{\theta=0},
     \quad K_{t,t^\prime}^{-1}[\varphi]=\frac{\delta^2
     S}{\delta\varphi(t)\delta\varphi(t^\prime)}\Big|_{\theta=0} .
        \end{split}
\end{gather}
Next  we integrate out the fast components $\theta$ and obtain the effective action for the slow components:
\begin{gather}
     S_{\rm eff}[\varphi]= S[\varphi]-\frac{1}{2}\int\frac{d\omega_1d\omega_2}{(2\pi)^2}
     b_{\omega_1}^\dagger
     K_{\omega_1\omega_2}b_{\omega_2}+\frac{i}{2}\tr\ln K\notag
     \\=S[\varphi]-S_I+S_{II} .  \label{action2}
\end{gather}
Here, frequencies\ $\omega_1,\omega_2$\ lie in the energy window\
$[\underline{\Lambda},\overline{\Lambda}],\
\underline{\Lambda}<\overline{\Lambda}$. The trace is understood to
be over the frequencies in the same window as well as in the Keldysh
space. High energy scale $\overline{\Lambda}$ in the AES-action is
naturally set by the first term in Eq.~\eqref{S_C_AES}:
$\overline{\Lambda} \sim g E_c$. Note that the linear in\
$\theta(t)$\ term in \eqref{action1} does not generally disappear.
But, as will be proven below, it is irrelevant since it leads to
$1/\overline{\Lambda}$ corrections.

Next  we perform the following decomposition
\begin{gather}
  K^{-1}[\varphi]=K^{-1}[0]+\Big(K^{-1}[\varphi]-K^{-1}[0]\Big)\notag \\
  \equiv
  K^{-1}[0]+\delta K^{-1}[\varphi]
\end{gather}
and treat the last term perturbatively. The operator\ $K^{-1}[0]$\
determines a fast field propagator. It corresponds to perturbative
Green function of the AES-action and follows from
Eqs.~\eqref{aes1}-\eqref{aes2}:
\begin{gather}
K^{R}(t,t^\prime) = - i \langle
\varphi_c(t)\varphi_q(t^\prime)\rangle, \,
K^{A}(t,t^\prime) = - i \langle \varphi_q(t)\varphi_c(t^\prime)\rangle ,\notag \\
K^{K}(t,t^\prime) = - i \langle
\varphi_c(t)\varphi_c(t^\prime)\rangle .
\end{gather}
In the leading order the Wigner transform of the perturbative Green
functions are given as
\begin{gather}
  \label{green1}
  \begin{split}
   K^{R}_{\omega}&=K^{A\dagger}_{\omega}=-\frac{4\pi i}{g}
   \left ( \int(F^d_\varepsilon-\frac{1}{g}\sum\limits_\alpha g_\alpha
   F^\alpha_{\varepsilon-\omega})\,d\varepsilon \right )^{-1} ,\\
   K^K_{\omega}&=2i\Imag K^{R}_{\omega} B_\omega ,
   \end{split}
\end{gather}
where we neglect all time derivatives with respect to slow time
since we are interested in high frequencies. The physical electron
distribution function is bound to have sign-function as its limit at
infinity\ $ F_\varepsilon\rightarrow\sgn(\varepsilon),\
\varepsilon\rightarrow\infty$. This yields the result
\begin{gather}
  \begin{split}
  K^{R}_{\omega}&=K^{A\dagger}_{p,\omega}=-\frac{2\pi i}{g}\frac{1}{\omega+\Delta
  Q \delta} , \\
   B_\omega&=\frac{\sum\limits_\alpha g_\alpha
    \int d\varepsilon(1-F^d_\varepsilon F^\alpha_{\varepsilon-\omega})}{2g (\omega+\Delta Q \delta)} , \\
\Delta Q&=\frac{\nu_d}{2}\sum_\alpha \int d\varepsilon \Bigl [
F_\varepsilon^d - \frac{g_\alpha }{g} F_\varepsilon^\alpha\Bigr ] .
  \end{split}
\end{gather}
In general, $\Delta Q$ does not vanish.
Next we find
\begin{gather}
  \begin{split}
   S_I&=\frac{1}{2}\int\frac{d\omega}{2\pi}\int dtdt^\prime b(t)^\dagger
     K[0]_{\omega}e^{-i\omega(t-t^\prime)}b(t^\prime)\\
     &-\frac{1}{2}\int\frac{d\omega_1d\omega_2}{(2\pi)^2}\int dtdt^\prime
     b^\dagger(t)e^{-i\omega_1(t-t_1)}
     K_{\omega_1}[0]\\
     &\times\delta K^{-1}[\varphi]_{t_1t_2}K_{\omega_2}[0]e^{-i\omega_2(t_2-t^\prime)}b(t^\prime) .
   \end{split}
\end{gather}
Performing integrations over\ fast frequencies $\omega,\ \omega_1,\
\omega_2$\ we see that the first integral is
$\sim1/((t-t^\prime)\overline{\Lambda})$\ and the second one is\
$\sim1/((t-t_1)(t^\prime-t_2)\overline{\Lambda}^2)$. Thus they are
irrelevant for RG-analysis. It means that only term $S_{II}$
contains logarithmic in $\overline{\Lambda}$\ corrections.

As usual we are interested in the first non-vanishing $\varphi$-dependent
correction:
\begin{gather}
  \label{trace}
  S_{II} \rightarrow\frac{i}{2}\tr\Big\{K[0]\delta
  K^{-1}[\varphi]\Big\} .
\end{gather}
Working out the trace in Eq.~\eqref{trace} we obtain
\begin{gather}
  \label{trace2}
  \begin{split}
 S_{II} &=-\frac{i}{2} \int dtdt^\prime \Bigg[\int\limits_{\overline{\Lambda}>|\omega|>
  \underline{\Lambda}}K^K_{\omega}(\tau)\frac{d\omega}{2\pi}\Bigg]  \\
  &\times \Big(\bar{X}_c(t)\bar{X}_q(t)\Big)
       \begin{pmatrix}
         0 &  \Pi^A(t,t^\prime)\\
        \Pi^R(t,t^\prime) &\ \Pi^K(t,t^\prime)
       \end{pmatrix}
       \begin{pmatrix}
             X_c(t^\prime)\\
             X_q(t^\prime)
         \end{pmatrix} .
   \end{split}
\end{gather}
Substituting it into~\eqref{action2} we see, that the
structure of the AES-action is restored. The only difference is  the
change of the coupling constant given by Eq.~\eqref{rg1}.
Finally, we mention that in the case of non-zero\ $\Delta Q$ Eq.~\eqref{Cond_O0}should be changed to $
  \omega_0 \sim\max\{\varepsilon_d,T_r,T_l, |\Delta Q|\delta\}$.


\section{Renormalization of the pseudo-fermion action \label{APPENDIX-RENORM}}

Here we provide details of the renormalization of the
pseudo-fermion action~\eqref{pseudo-action1} which are used in Sec.~\ref{RELAX-STRONG}.

\subsection{Renormalization of $Z$,\ $\Delta$,\  and $g$.}

\begin{figure}[t]
  \includegraphics[width=55mm]{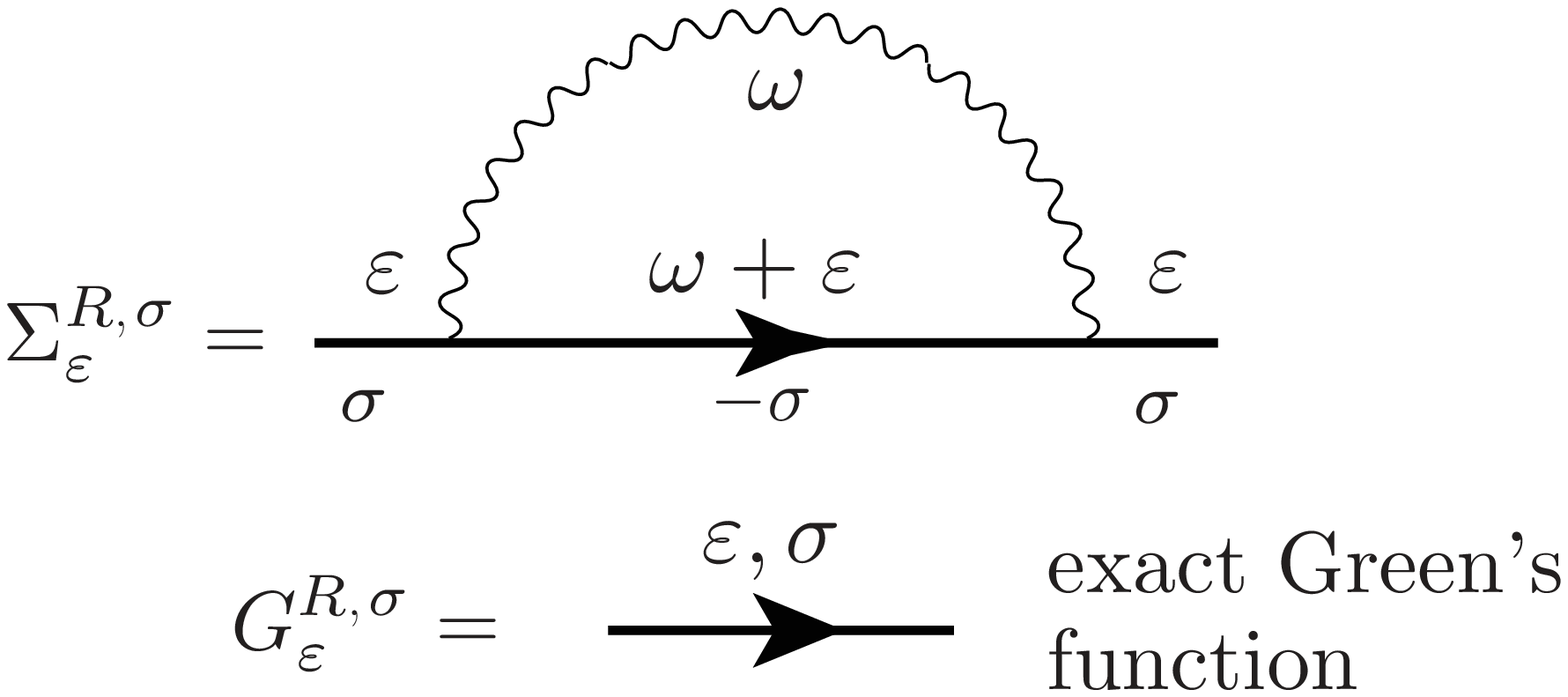}
  \caption{\label{figure7}
   The Dyson equation for pseudo-fermion self-energy.
          }
\end{figure}

The exact pseudo-fermion Green's function can be written as
\begin{gather}
   \overline{G}^{R}_{\varepsilon,\sigma}=\frac{1}{\varepsilon-\xi_\sigma-\Sigma^{R,\sigma}_\varepsilon}.
\end{gather}
Here,\ $\xi_\sigma=\Delta\sigma/2-\eta$. To write it in the
renormalized form~\eqref{green1_0} we redefine the theory's
constants and write down the standard relations defining the Green's
function scaling $Z$, the renormalized gap $\bar{\Delta}$\ and the
Green's function width $\Gamma^{\sigma}_\varepsilon$\ respectively:
\begin{gather}
   Z=\Big(1-\partial_\varepsilon\Re\Sigma^{R,\sigma}\big|_{\varepsilon=\bar{\xi}_\sigma}\Big)^{-1},\label{a0}\\
   \bar{\xi}_\sigma=\xi_\sigma+\Re\Sigma^{R,\sigma}\big|_{\varepsilon=\bar{\xi}_\sigma},\label{a1}\\
   i\bar{g}\Gamma^\sigma_{\varepsilon}=-iZ\Im\Sigma^{R,\sigma}_{\varepsilon}.\label{a2}
\end{gather}
To find the scaling\ $Z$ and relate $\bar{g}$, and $\bar{\Delta}$ to
their bare counterparts we solve the one-loop Dyson equation for the
self-energy presented in Fig.~\ref{figure7}. With the help
of~\eqref{self-energy1} we find
\begin{gather}
\label{delta2}
  \begin{split}
 \Re\Sigma^{R,\sigma}_\varepsilon&=\frac{g}{4\pi}\int\frac{d\omega}{2\pi}\omega B_\omega
 Z(\omega)\\
 &\times\Re
 \frac{1}{\varepsilon+\omega\sigma-\bar{\xi}_{-\sigma}
 -iZ\Im\Sigma^{-\sigma,R}_{\varepsilon+\omega\sigma}}
   \end{split}
\end{gather}
It is important to understand that scaling parameter\ $Z$\ cannot be
put before the sign of an integral. Generally it is cut-off
dependent and contains the factor $\ln(\Lambda/\omega_0)$, where
$\Lambda$ is an ultraviolet cut-off of the theory ($E_c$\ in our
case) while\ $\omega_0$ is a characteristic scale of the Green's
function entering the integrand. To determine $\omega_0$ we notice
that the integral in~\eqref{delta2} diverges, being determined by
the behavior of the integrand in the large\ $\omega$ limit. That is
why the characteristic scale of the Green's function
entering~\eqref{delta2} is its running frequency:
$\omega_0\sim\omega$. Solving Eqs.~\eqref{delta2} and~\eqref{a0}
with logarithmic accuracy we obtain
\begin{gather}
  \label{sol1}
   \frac{1}{Z^2(\Lambda)}\frac{\partial Z(\Lambda)}{\partial\Lambda}=\frac{g}{4\pi^2}\frac{B(\Lambda)Z(\Lambda)}
   {\Lambda}
\end{gather}
Integrating~\eqref{sol1} in the limits\ $[\omega_0,E_c]$ we
recover~\eqref{green1_1} in complete analogy with equilibrium case.
Thus, the renormalization procedure is outlined and the rest of
formulae~\eqref{green1_0}-\eqref{green1_2} are obtained in a similar
fashion.

\subsection{Callan-Symanzik equation for $\langle\mathcal{N}\rangle_{pf}$.}

The anomalous dimension\ $\gamma$\ of $\langle\mathcal{N}\rangle$\
is introduced as
\begin{gather}
  \label{RG1}
  Z^\gamma \overline{\langle\mathcal{N}\rangle}_{pf} (\bar{\Delta},\bar{g})=\langle\mathcal{N}
  \rangle_{pf} (\Delta,g,\Lambda).
\end{gather}
To extract $\gamma$ we write down the corresponding Callan-Symanzik equation
for:\
$\langle\mathcal{N}\rangle_{pf}(\Delta,g,\Lambda)=\sum_\sigma\langle\bar{\psi}_\sigma\psi_\sigma\rangle$.
The tree-level $\langle\mathcal{N}\rangle_{pf}(\Delta,g,\Lambda)$\ is given by Eq.~\eqref{density1}.
Following general strategy we write the corresponding
Callan-Symanzik equation for the function\
$\langle\mathcal{N}\rangle_{pf}(\Delta,g,\Lambda)$\ in the form:
\begin{gather}
  \label{CS-equation}
  \Big(\frac{\partial}{\partial\ln \Lambda}+\beta_g\frac{\partial}{\partial g}+
  \beta_\Delta\frac{\partial}{\partial\Delta}+\gamma
  \frac{g}{4\pi^2}\Big)\langle\mathcal{N}\rangle(g,\Delta,\Lambda)=0.
\end{gather}
where the corresponding\ $\beta$\ - functions are easily seen from
Eq.~\eqref{green1_2}:
\begin{gather}
  \label{RG}
  \begin{split}
   \beta_g=\frac{g^2}{2\pi^2},\quad
   \beta_\Delta=\frac{g\Delta}{2\pi^2}.
   \end{split}
\end{gather}
The term with\ $\beta_g$\ always contains extra\ $g$\ and can be
dropped in the leading order.

\begin{figure}[t]
  \includegraphics[width=30mm]{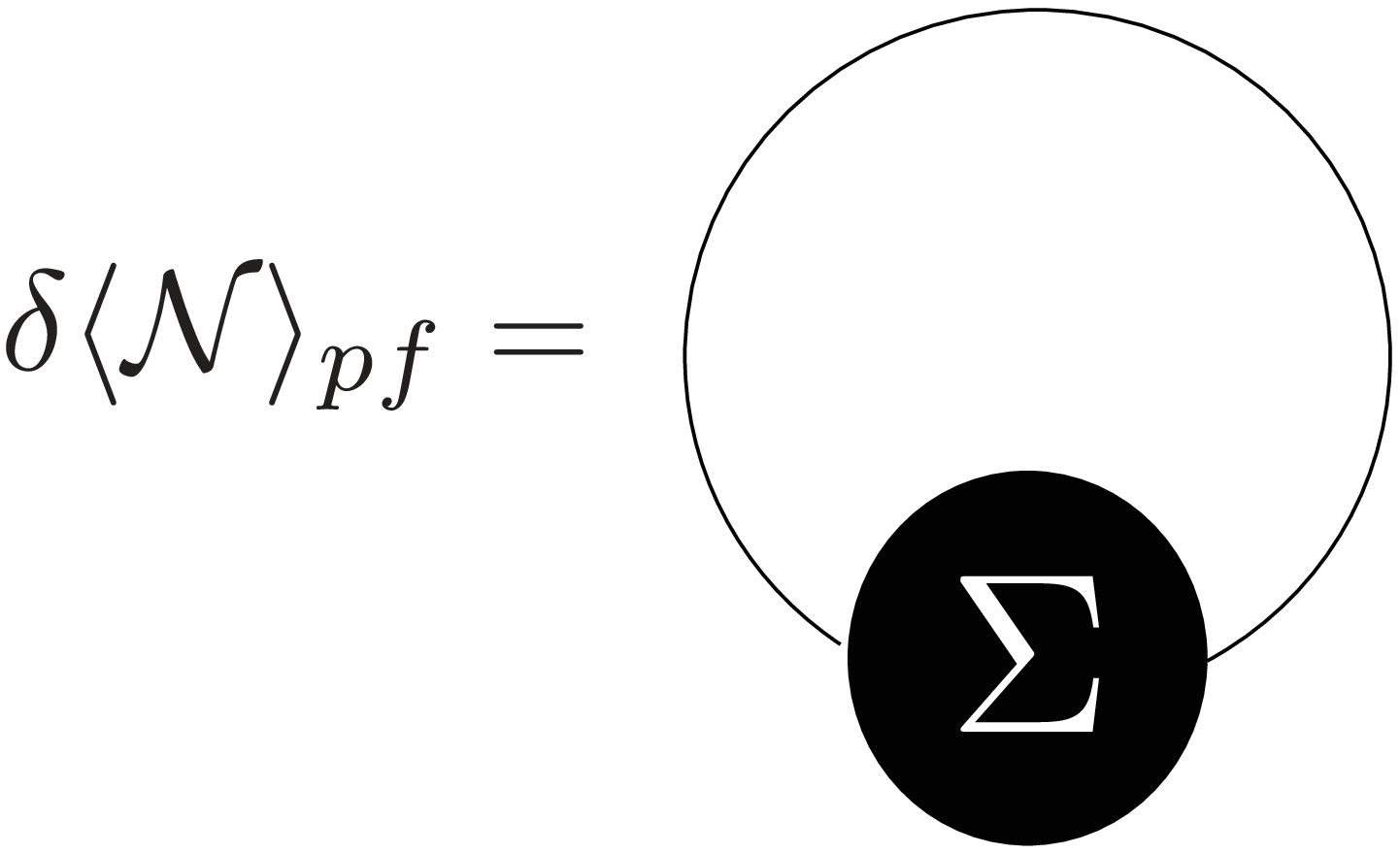}
  \caption{
   Correction to pseudo-fermion particle number\ $\langle\mathcal{N}\rangle_{pf}$.
          }\label{figure8}
\end{figure}

To find\ $\gamma$\ we need to find $\langle\mathcal{N}\rangle_{pf}$\
in the next to the tree-level order. The diagram representing the
correction to pseudo-fermion particle number is presented in
Fig.~\ref{figure8}. Calculating with logarithmic accuracy and using
extensively the fact that ${\cal B}_\omega\rightarrow\sgn\omega$ at
large $\omega$ we obtain
\begin{gather}
  \label{RG3}
  \begin{split}
  \langle\mathcal{N}\rangle_{pf}(\Delta,g,\Lambda)&=1-\frac{{\cal F}^++{\cal F}^-}{2}\\
  &-\frac{g}{8\pi^2}\Delta(\partial_\eta {\cal F}^+-\partial_\eta {\cal F}^-)
    \int_{\omega_0}^{\Lambda}\frac{d\omega}{\omega}B(\omega).
  \end{split}
\end{gather}
 Plugging\ Eq. \eqref{RG3}) into Eq. \eqref{CS-equation}) we find that
$ \gamma=0$ that proves Eq.~\eqref{density-scaling}.

\begin{thebibliography}{100}
\bibitem{Footnote1} Formally, the integral in Eq.~\eqref{KEsim1} for $\mathcal{G}(\tau)$
diverge at $\omega \to 0$. Howerver, this divergence is due to the presence of unphysical
zero mode of the boson field $\varphi$ in the AES-action \protect{\eqref{S_AES}}. Exclusion
of this zero mode leads to substitution of $\lim\limits_{\epsilon\to 0}
\omega^3/(\omega^2+\epsilon^2)^2$ for $1/\omega$ in Eqs.~\eqref{KEsim1}.
Obviously, it does not affect the form of Eq.~\protect{\eqref{diff1}}
\end{thebibliography}
\end{document}